\documentclass{aa}
\usepackage{graphicx}
\usepackage{natbib}
\usepackage{supertabular}
\bibpunct{(}{)}{;}{a}{}{,}

\newcommand{\bra}{Br$\alpha$}

\newcommand{\hei}{\ion{He}{i}}
\newcommand{\heii}{\ion{He}{ii}}
\newcommand{\niii}{\ion{N}{iii}}
\newcommand{\civ}{\ion{C}{iv}}

\newcommand{\HII}{\ion{H}{ii}}
\newcommand{\brg}{Br$\gamma$}
\newcommand{\micron}{$\mu$m}
\newcommand{\kms}{km~s$^{-1}$}

\begin{document}

   \title{\emph{VLT} $K$-band spectroscopy of massive
   stars deeply embedded in IRAS sources with  UCHII colours.
   \thanks{Based on observations collected at the European Southern 
   Observatory at La Silla and Paranal, Chile (ESO programs 64.H-0425
   and 65.H-0602)}
   \thanks{Table 6 is only available in electronic form at CDS.}}

   \subtitle{}

  \author{A.~Bik \inst{1,2}
          \and L.~Kaper \inst{1}
	  \and M.~M.~Hanson \inst{3}
	  \and M.~Smits \inst{1}
          }

   \offprints{A.~Bik (abik@eso.org)}

   \institute{Astronomical Institute ``Anton Pannekoek'',
           University of Amsterdam, Kruislaan 403, 1098 SJ Amsterdam,
	   The Netherlands
	  \and European Southern Observatory, Karl-Schwarzschild
           Strasse 2, Garching-bei-M\"unchen, D-85748, Germany
         \and University of Cincinnati, Cincinnati, OH 45221-0011, U.S.A.
             }

   \date{}

    \authorrunning{A.~Bik et al.}
    \titlerunning{$K$-band spectroscopy of massive stars in IRAS sources
    with UCHII colours}

   \abstract{We have obtained high resolution ($R$ = 10,000)
   $K$-band spectra of candidate young massive stars deeply embedded
   in (ultra-) compact \HII\ regions (UCHIIs). These objects were
   selected from a near-infrared survey of 44 fields centered on IRAS
   sources with UCHII colours. Often, the near-infrared counterpart of
   the IRAS source is a young embedded cluster hosting massive
   stars. In these clusters, three types of objects are identified. The
   first type (38 objects) consists of ``naked'' OB stars whose
   $K$-band spectra are dominated by photospheric emission. We
   classify the $K$-band spectra of the OB-type cluster members using
   near-infrared classification criteria. A few of them have a very
   early (O3-O4~V) spectral type, consistent with a young age of the
   embedded clusters.  The spectral classification provides an
   important constraint on the distance to the embedded cluster.
 The ionising power of the population thus derived is compared to the
   information obtained from the infrared and radio flux of these
   sources. In most cases these two different determinations of the
   ionising flux are consistent, from which we conclude that we have
   identified the ionising star(s) in about 50 \% of the embedded
   clusters.  The second type (7 objects) are point sources
   associated with UCHII radio emission, that exhibit nebular emission
   lines in the near-infrared. Six of the objects in this group
   produce \hei\ emission indicative of an embedded O-type star. These
   objects are more embedded than the OB stars and probably do not
   dominate the infrared flux as measured by IRAS. They may emit the
   bulk of their reprocessed UV radiation at mm wavelengths. The
   third type (20 objects) is characterised by broad
   (100--200~\kms) \brg\ emission and no photospheric absorption
   profiles. \citet{Brgspec04} show that these objects are massive YSO
   candidates surrounded by dense circumstellar disks.

 \keywords{Stars: early-type, formation, distances, ISM: HII regions,
 Infrared: stars} }

   \maketitle
%

\section{Introduction}\label{sec:introduction}

The recent advent of high-quality, near-infrared instrumentation has
opened up a new window on the birth sites of massive stars. The
formation timescale of a massive star ($\geq 10$~M$_{\odot}$) is short,
of the order of 10,000~yr, so that newly formed stars are still deeply
embedded inside their parental molecular cloud. 

Observational evidence
indicates that at this early stage in the evolution the ``hot core''
contains an extended circumstellar disk \citep[size $\sim 1000$~AU;
e.g.,][]{Beltran04,Chini04} through which a vast amount of material
would be able to lose its angular momentum and accrete onto the
protostar. Before arriving on the main sequence, the young
massive star already will produce a strong UV radiation field, ionising its
surroundings. At this stage the object will become detectable at
far-infrared and radio wavelengths through the heated dust and
recombination of hydrogen in the expanding hyper- or ultra-compact
\HII\ region \citep[e.g.][]{Churchwell02}. The
same UV radiation field will impact on the extended circumstellar disk
and start to destroy it \citep{Hollenbach94}. The disk
destruction timescale is short too ($\sim 100\,000$~yr), so that it
becomes an observational challenge to detect and measure any remnants
of the formation process of massive stars.

At near-infrared wavelengths one is able to penetrate the obscuring
gas and dust in the natal cloud and may detect the
photospheric and circumstellar emission of the young massive star. We
have carried out an extensive near-infrared survey of fields centered
on IRAS sources with the characteristic colours of ultra-compact \HII\
regions \citep[UCHIIs,][]{WoodIRAS89} with SOFI on the ESO {\it New
Technology Telescope} \citep{Kaper04}.

 The main result in this survey is that in all 44 regions but one
a near-infrared source associated with \brg\ emission is found.  In 75
\% of the fields this near-infrared emission corresponds to an
embedded massive stellar cluster ($A_{V} \sim 10-20$~magnitudes). Only
a small fraction of the near-infrared sources have a size typical of
UCHII regions i.e. 0.1 pc.  Besides the embedded clusters, we have
detected the near-infrared counterpart of a dozen ultra-compact radio
sources (i.e. in 40~\% of the fields in which an ultra-compact radio
source is located).

This paper concentrates on the spectroscopic follow-up of the sources
selected from our near-infrared NTT survey to confirm
their massive-star nature and to search for observational signatures
that may reveal information on the formation process. $K$-band spectra
have been obtained of candidate young massive stars identified in our
survey with ISAAC mounted on the ESO {\it Very Large Telescope}, based
on their $K$-band magnitude and $J-K$ colour. These candidate massive
stars are  members of the embedded cluster or, in some cases, the
potential ionising stars of UCHIIs.

The $K$-band window is well suited to perform the spectral
classification of OB-type stars \citep{Hanson96}, especially when
interstellar extinction becomes important. However, in order to
retrieve the photospheric spectrum of an embedded massive star, one
has to correct for the strong nebular emission produced by the
surrounding \HII\ region. As OB-type stars mainly include lines
of hydrogen and helium in their spectra, many photospheric lines will
be affected by nebular emission. When using ISAAC with a 0.3~arcsec
slit, the spectral resolution ($R \simeq 10\,000$) becomes sufficient
to resolve the broad photospheric lines and to
discriminate between the nebular and stellar contribution. The
feasibility of this approach has been demonstrated by
\cite{WatsonSpectra97} who retrieved the photospheric spectrum of the
O5--O6.5~V star ionising the famous cometary UCHII
\object{G29.96-0.02}  \citep[see also][]{MartinG2903}. In the mean time
the photospheric spectra have been registered of massive stars
embedded in a handful of UCHIIs \citep[e.g.][]{Hanson02}.

Through the direct measurement of the spectral properties of the
deeply embedded, young  massive stars, a comparison can be made with
estimates based on the far-infrared and radio emission produced by the
UCHII c.q. embedded cluster. Following this procedure, \citet{Flame03}
were able to identify the long-sought-for ionising star of the Flame
Nebula (\object{NGC 2024}). As a by-product of the spectral
classification, the amount of interstellar extinction $A_{V}$ can be
determined as the intrinsic $J-K$ colour of OB stars falls within a
narrow range. A relation between spectral type and luminosity then
helps to constrain the distance to the embedded cluster, providing
important information on the distribution of the embedded massive star
population in the Milky Way. Furthermore, one can investigate whether
these young massive stars differ from the OB-star field population,
e.g. with respect to their wind and rotation properties, and whether
they show any signs that reveal information on the formation process.


The paper is organised as follows; the next section describes the
VLT/ISAAC spectroscopic observations and the applied data reduction
procedure. In Sect. \ref{sec:spectra} we introduce the $K$-band
spectra  and divide them into different classes. We also outline
  the spectral classification method. Using the
spectroscopic parallax, the distances to the OB stars are estimated
and compared to literature values.  Sect. \ref{sec:uchiicp} includes
the $K$-band spectra of the near-infrared counterparts of the UCHII
regions detected at radio frequencies.  In Sect. \ref{sec:regions} the
luminosity of the OB star population thus identified is compared to
that inferred from the infrared and radio
measurements. Sect. \ref{sec:discussion} discusses the properties of
the young OB stars with respect to those measured in the field
population. The last section summarises our conclusions.

\section{Observations}\label{sec:observations}

$K$-band spectra were taken of point sources in 31 regions from
the sample of 44 IRAS sources with UCHII
colours presented in \citet{Kaper04}.  The point sources were
selected based on their position in the colour-magnitude diagram. The
brightest and most reddened sources were selected for follow-up
spectroscopy.

Medium-resolution, long-slit ($120\arcsec$) $K$-band spectra were
obtained with ISAAC mounted on Antu (UT1) of ESO's \emph{Very Large
Telescope} (VLT), Paranal, Chile. The observed targets and the
observing log are given in Table \ref{tab:obslog}.  The observations
were carried out during a visitor mode run (by LK and MMH) on March
18--20, 2000, and a service mode run in the period between May 18 and
July 16, 2000.  A slit width of $0.3\arcsec$ was used, resulting in a
spectral resolving power $R = 10,000$. The slit position was
chosen such that at least two sources were positioned on the slit. The
observing conditions at Paranal Observatory were excellent: humidity
less than 10\%\ and seeing ranging from $0.5-1.0$\arcsec.

The central wavelength was chosen such that the most important lines
needed to classify the OB stars in the $K$-band are covered (see
Sect. \ref{sec:classification}). The spectra taken in visitor mode
(the objects with R.A. between 5 and 15 hour) have a central
wavelength of 2.129 \micron\ and  range from 2.069 to 2.189 \micron.
This setting should have covered the \heii\ line at 2.1885 \micron; it turned
out, however, that the last 20 pixels were vignetted and not
usable. In the service mode run (sources with R.A. between 16 and 19
hour) the central wavelength was shifted to 2.134 \micron\ in order
to cover the \heii\ line.

The electrical ghosts and bias were removed from the frames before
flatfielding.  Despite the use of flatfields taken during the night,
some service mode flatfields have a fringe pattern different from
that of the science frames and thus introduce fringes in the final
spectrum.  These flatfields were not used and replaced by flatfields
obtained in other nights which cause less fringing; in some spectra,
however, remnants of fringes are still visible. The wavelength
calibration of the spectra is performed using arc spectra with the
IRAF task \emph{identify}. The typical error of the wavelength
solution is 3 \kms.

In order to correct for the sky background, the object was ``nodded''
between two positions on the slit (A and B) such that the background
emission registered at position B (when the source is at position A)
is subtracted from the source plus sky background observation at
position B in the next frame, and vice versa. The offsets were
such that the two stars positioned on the slit are not overlapping. In
some sources, the nebular lines (e.g. \brg) are spatially extended and will
overlap resulting in artifacts in the sky subtracted image. This
can mimic a narrow absorption feature superimposed on the stellar \brg\ line.

Telluric absorption lines were removed using stars of spectral type
B8~V -- A2V, observed under identical sky conditions.  Before removing
the telluric lines, the only photospheric line (\brg) in the spectrum
of the A star needs to be divided out. It turns out that the best
result is achieved when first the telluric features are removed from
the $K$-band spectrum of the telluric standard using a high-resolution
telluric spectrum (obtained at NSO/Kitt Peak). This spectrum is taken
under very different sky conditions, so numerous spectral remnants remain
visible in the corrected standard star spectrum. Without this
``first-order'' telluric-line correction, a proper fit of \brg\ cannot
be obtained. The \brg\ line is fitted by a combination of two
exponential functions. The error on the resulting \brg\ equivalent
width (EW) of our target star is about 5~\%. After the removal of the
\brg\ line of the A star, the telluric lines are removed by taking the
ratio of the target spectrum and the telluric spectrum.  This is done
using the IRAF task \emph{telluric}, which allows for a shift in
wavelength and a scaling in line strength, yielding a more accurate
fit.  The task uses a cross-correlation procedure to determine the
optimal shift in wavelength and the scaling factor in line strength,
which can be changed interactively. The shifts are usually a few
tenths of a pixel; also the scaling factors are modest ($\sim 10$~\%).

\section{$K$-band spectra of young OB stars}\label{sec:spectra}

\begin{figure}
\begin{center}
 \includegraphics[width=\columnwidth]{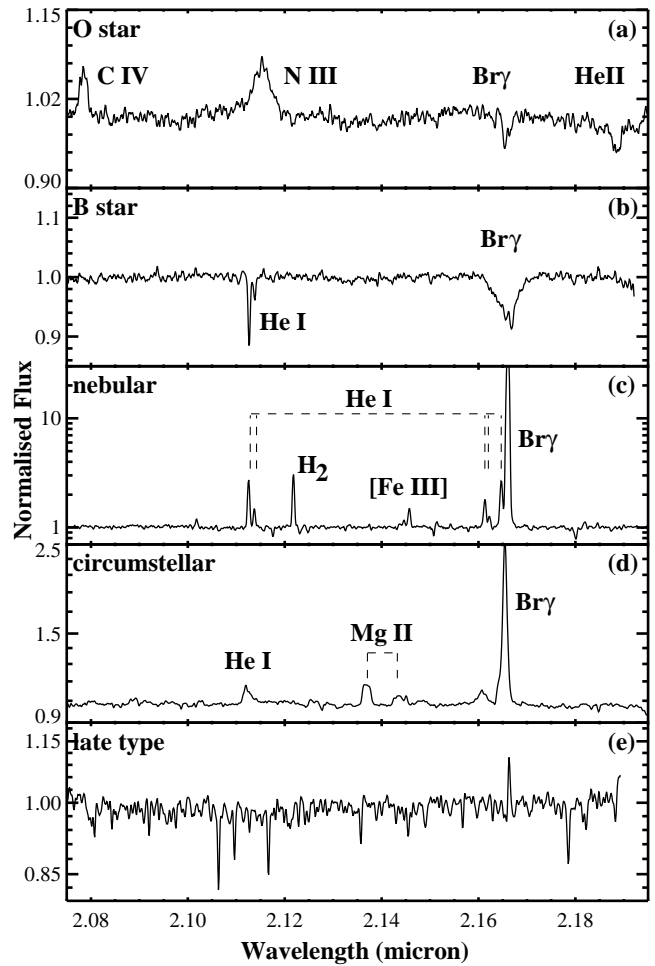}
  \caption{$K$-band spectra of the different classes of objects found
  in our spectroscopic survey. Panels (a) and (b) show the
  photospheric $K$-band spectra of O and B stars which is the main
  focus of this paper. Panel (c) shows a spectrum dominated by nebular
  emission from the UCHII region. The emission lines are narrow and
  not resolved. This in contrast to the spectrum displayed in panel
  (d) where the emission lines are resolved. This object
  shows characteristics of a massive YSO. Panel (e) displays a
  spectrum of a late type (fore-/background) star. The spectrum is
  dominated by absorption lines. The narrow \brg\ emission line
  originates in the \HII\ region. Note that panels (c) and (d) are plotted
  on a logaritmic scale. \label{fig:example}}
\end{center}
\end{figure}

The $K$-band spectra can be divided in different classes (see
Fig. \ref{fig:example} for an example of the different spectra). 
Forty objects turn out to be late-type (fore- or
background) stars, with $K$-band spectra dominated by many absorption
lines (Fig. \ref{fig:example}e). These stars are a natural
by-pro\-duct of our selection criteria. They emit the bulk of their
radiation in the (near) infrared and have a red intrinsic colour. The
other three classes are different types of objects connected with
star-forming regions.

Thirty-eight objects show the spectral features indicative of OB
stars and are the subject of this study
(Fig. \ref{fig:example}a,b). In the following, we will describe the
classification method introduced by \citet{Hanson96}. The spectra are
presented and classified, distances are estimated based on the
$K$-band spectral type (spectroscopic parallax). The second type of
objects discussed in this paper are the near-infrared
counterparts of the UCHII regions with nebular $K$-band spectra
(Fig. \ref{fig:example}c).

Another 20 objects show broad, spectrally resolved \brg\ emission
(Fig. \ref{fig:example}d) and are presented in \citet{Brgspec04}. 
These emission-line objects are classified as massive Young Stellar
Objects (YSOs). Towards some of these objects CO first-overtone
emission is detected, originating from warm (2000-4000 K) and dense
($10^{10}$ cm$^{-3}$) material. \citet{COletter04} show that the CO
emission in these objects is coming from a keplerian rotating disk
within 5 AU from the central star. These objects are very likely
surrounded by the remnant of a accretion disk.  Photo-evaporation
models show that the outer parts of the accretion disk get
photo-evaporated on a very short timescale, while the inner parts of
the disk remain present around the recently formed star for a longer
time  \citep{Hollenbach94}.

\subsection{Classification scheme}\label{sec:classification}

\begin{figure}
\begin{center}
 \includegraphics[width=\columnwidth]{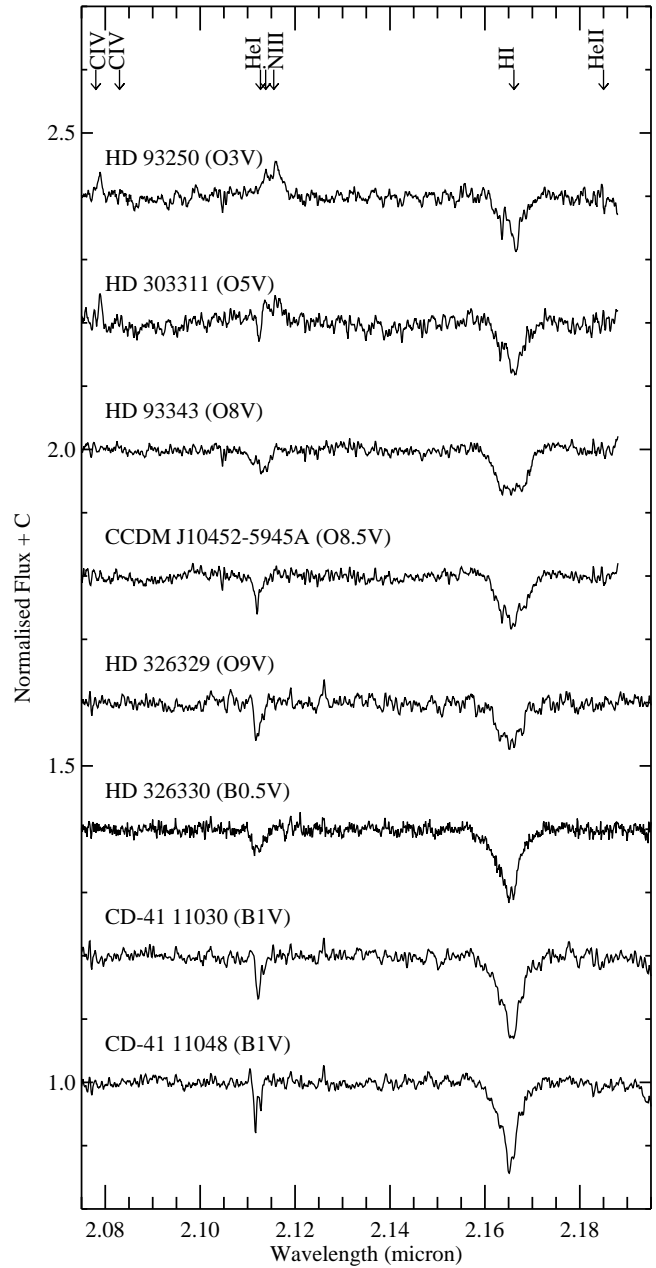}
  \caption{$K$-band spectra of optically visible OB stars in
  \object{Tr 16} and \object{NGC 6231}.  Application of the
  classification criteria introduced by \citet{Hanson96} yields
  spectral types consistent with the MK classification. The higher
  spectral resolution results in the detection of the \hei\
  line in O5~V stars, which is not seen in the low
  resolution spectra of \citep{Hanson96}.}
   \label{fig:refstars}
\end{center}
\end{figure}

\citet{Hanson96} constructed a $K$-band classification scheme
based on the relative line strength in spectra of OB stars with known
(optical) MK spectral type. The stars were independently classified, and the
derived $K$-band classes compared with MK spectral types.

The lines covered by the $K$-band spectra of massive stars
are a hydrogen line (\brg\, 2.16 \micron), three \hei\ lines, one at
2.058 \micron\ (not in our spectral setting), and  two at
2.112 and 2.113 \micron. Also a \heii\ line is present at 2.185
\micron. The other lines detected in O star spectra are a complex of
\niii\ lines at 2.115 \micron\ and three \civ\ lines with the two
strongest at 2.069 and 2.078 \micron\ and a very weak line at 2.083
\micron.

The \brg\ line is present in both the O and B type stars, and
increasing in strength from the early O toward late B/early A. The
\heii\ line strength is decreasing toward later spectral type and the
line disappears when the spectral type becomes later than O7/O8. The
\civ\ (emission) lines are only visible for a small range in effective
temperature. \citet{Hanson96} find that stars with spectral type
between O5 and O6.5, and down to O8 in supergiants, have \civ\ lines in their
spectrum, making it a sensitive temperature diagnostic. The
identification of the emission complex  at 2.1155 micron is still
uncertain, but is assumed to be \niii. The line is
visible in stars with spectral type between O3 and O8. The \hei\ lines
at 2.112/2.113 \micron\ start to become visible in O7-O8 stars and
remain present till B3 in dwarfs and as late as B8 or B9 in
supergiants.

The classification scheme developed by \citet{Hanson96} is summarised
in Table \ref{tab:scheme}. The classification of OB stars in the
$K$-band is hampered by the limited number of lines as well as the few spectra
available. The $K$-band classification is therefore not
as accurate as the optical classification. Especially for the B
stars, where only two lines are present (\hei\ and \brg), the
classification becomes less reliable.

\begin{table*}
\begin{center}
\begin{tabular}{llllllp{1cm}} \hline\hline
\multicolumn{1}{c}{$K$-band} &\multicolumn{1}{c}{MK}       &\multicolumn{1}{c}{\civ}  &\multicolumn{1}{c}{\niii} & \multicolumn{1}{c}{\hei}  &\multicolumn{1}{c}{\ion{H}{i}(\brg)} &\multicolumn{1}{c}{\heii}  \\ 
\multicolumn{1}{c}{Spec. type}&\multicolumn{1}{c}{Spec. type} &\multicolumn{1}{c}{2.078 \micron}  &\multicolumn{1}{c}{2.115 \micron} & \multicolumn{1}{c}{2.113 \micron}  &\multicolumn{1}{c}{2.166 \micron} &\multicolumn{1}{c}{2.185 \micron}  \\ 

&     &\multicolumn{1}{c}{(\AA)}  &\multicolumn{1}{c}{(\AA)} & \multicolumn{1}{c}{(\AA)}  &\multicolumn{1}{c}{(\AA)} &\multicolumn{1}{c}{(\AA)}  \\ \hline
\noalign{\smallskip}
\multicolumn{7}{c}{Dwarfs and Giants}                                                  \\ \hline \hline
kO3--O4           & O3V--O4V             & np         & em         & np       & $\leq 4(5)$   & abs   \\ 
kO5--O6           & O5V--O6.5V           & em         & em         & np       & $\leq 4(5)$   & abs   \\ 
kO7--O8           & O7V--O8V             & w/np       & em         & abs      & $\leq 4(5)$  & w,abs \\
kO9--B1           & O8V--B1V             & np         & np         & abs      & $\leq 4(5)$  & np    \\
kB2--B3           & B1V--B2.5V           & np         & np         & abs      & $4(5)-8$     & np    \\
kB4--B7           & B3V--B7V             & np         & np         & np       & $4(5)-8$     & np    \\
kB8--A3           & B8V--A3V             & np         & np         & np       & $> 8$        & np    \\\hline 
\noalign{\smallskip}			                             			 
\multicolumn{7}{c}{Supergiants}                                                \\ \hline \hline
kO3--O4b          & O3I--O4I              & np        & np         & np       & w/em     & em,abs \\
kO5--O6b          & O5I--O6.5I            & em        & em         & np       & w/em     & abs  \\
kO7--O8b          & O7I--O8I              & w/np      & em         & abs      & w/em     & w,abs \\
kOBb              & O9I--B3I              & np        & np         & abs      & em       & np   \\ \hline
\end{tabular}
\caption{Spectral classification of OB stars as proposed by
  \citet{Hanson96}. \emph{abs}: in absorption, \emph{em}: in emission,
  \emph{w}: weak, \emph{np}: not present. For the lines not detected
  in the low-resolution spectra, upper limits are used; for \niii: 0.2
  \AA\ and for the other lines 0.3 \AA\ is used as upper limit. For
  the \brg\ criterium, between brackets the modification as discussed
  in the text is indicated. The $K$-band spectral type (column 1) is
  based on the presence of the lines listed in this Table and is
  derived independently of the corresponding MK spectral type.  In the
  $K$-band spectra of early-O stars the \niii\ and \civ\ lines are
  important diagnostics. At late-O the \hei\ line starts to become
  important, while for B stars, the main diagnostic is the strength of
  the \brg\ line.\label{tab:scheme}}
\end{center}
\end{table*}

As the spectral resolution and signal-to-noise ratio of the spectra
presented in this paper are different from the data used by
\citet{Hanson96}, care should be taken in applying the classification
of \citet{Hanson96} right away.  To be able to compare our
observations with the low-resolution spectra of \citet{Hanson96} we
observed spectra of optically visible OB stars with VLT/ISAAC.  The
optically visible OB stars are located in the \object{Tr 16} cluster
in the Car OB2 association \citep{The80} and \object{NGC 6231} in the
\object{Sco OB1} association \citep{Sung98}. The spectra are shown in
Fig. \ref{fig:refstars}, plotted in order of optical MK-spectral
type. The spectra of the O stars in \object{Car OB2} were taken during
the visitor mode run, when the \heii\ line was not well covered. In
the $K$-band spectra of the stars in \object{NGC 6231}, the \heii\
line is not present as the stars are of later spectral type.

Applying the classification criteria developed by \citet{Hanson96} to
the spectra of the ``reference stars'' does for some of the stars not
result in a classification consistent with their MK spectral type. The
spectra of the classification stars in \citet{Hanson96} have a
resolving power varying between $R =800$ and $R=1100$, with a few stars
taken at a resolution of 3000. While the spectra presented in this
paper are taken with a spectral resolution of 10,000 and in the most
cases higher signal-to-noise ratio.  The difference in spectral
resolution and signal-to-noise between our spectra and those used by
\citet{Hanson96} results in slightly different classification
criteria.

\begin{table*}
\begin{center}
\begin{tabular}{lllllll} \hline\hline
\multicolumn{1}{c}{Object} &\multicolumn{1}{c}{\civ}  &\multicolumn{1}{c}{\niii} & \multicolumn{1}{c}{\hei}  &\multicolumn{1}{c}{\ion{H}{i}(\brg)} &\multicolumn{1}{c}{$K$-band} &\multicolumn{1}{c}{Optical}\\ 
&\multicolumn{1}{c}{(\AA)}  &\multicolumn{1}{c}{(\AA)} & \multicolumn{1}{c}{(\AA)}  &\multicolumn{1}{c}{(\AA)} &\multicolumn{1}{c}{Sp.Type}&\multicolumn{1}{c}{Sp.Type} \\ \hline
\noalign{\smallskip}
\object{HD 93250} (O3V)              & $-0.49 \pm 0.1$  & $-2.11 \pm 0.35$   & --                & $3.67 \pm 0.3$   & KO5--O6  & O5V--O6.5V   \\ 
\object{HD 303311} (O5V)             & $-0.58 \pm 0.1$  & $-1.17 \pm 0.4$    & $0.28 \pm 0.07$   & $4.26 \pm 0.47$  & KO5--O6  & O5V--O6.5V  \\ 
\object{HD 93343} (O8V)              & --               & --                 & $1.24 \pm 0.27$   & $5.04 \pm 0.4$   & kO9--B1  & O8V--B0V/B1V  \\
\object{CCDM J10452-5945A} (O8.5V)  & --               & --                 & $0.82 \pm 0.12$   & $ 4.63 \pm 0.33$ & kO9--B1  & O8V--B0V/B1V  \\
\object{HD 326329} (O9V)             & --               & --                 & $0.94 \pm 0.14$   & $ 3.54 \pm 0.43$ & kO9--B1  & O8V--B0V/B1V  \\
\object{HD 32633} (B05V)              & --               & --                 & $0.99 \pm 0.21$   & $ 6.31 \pm 0.46$ & kB2--B3  & B1V--B2.5V  \\
\object{CD-41 11030}(B1V)          & --               & --                 & $0.94 \pm 0.11$   & $ 7.06 \pm 0.39$ & kB2--B3  & B1V--B2.5V  \\  
\object{CD-41 11048} (B1V)          & --               & --                 & $0.81 \pm 0.10$   & $ 6.54 \pm 0.32$ & kB2--B3  & B1V--B2.5V  \\\hline 
\end{tabular}

\caption{EW measurements and $K$-band classification of the optically
  visible reference stars in \object{Tr16} and \object{NGC 6231} (Fig. \ref{fig:refstars}). Column 1 gives
  the name and the spectral type as provided by Simbad. Column 6 and 7 list the
  $K$-band spectral type applying the classification scheme of
  \citet{Hanson96} and the corresponding optical MK classification.\label{tab:refstars}}
\end{center}
\end{table*}

In a low-resolution spectrum the \hei\ 2.112/2.113 lines are not
resolved and are observed as one line.  In our spectra, these lines are
resolved and can provide information on the shape of the line. The
higher spectral resolution and the higher signal-to-noise ratio of our
spectra enables the  detection of  lines with lower EW than could be detected
by \citet{Hanson96}.  An illustration of this is the presence of the
\hei\ line in the spectrum of \object{HD 303311}. The star is
optically classified as O5V, which, according to classification
of \citet{Hanson96}, does not show the \hei\ line in the spectrum. However,
the \hei\ line is detected in our high-resolution spectrum. When
smoothed to a resolution of $R=1100$, the \hei\ line disappears.  This
\hei\ line is also visible in the newly taken higher resolution
spectra of O5V-O6V stars (Hanson et al., in prep.). Therefore, upper
limits reflecting the smallest EW detected by \citet{Hanson96} are
used in the classification scheme for lines that were not detected by
\citet{Hanson96}. The EW of the \hei\ line in the spectrum of
\object{HD 303311} falls below the adopted upper limit for the \hei\
line (0.3 \AA).

\begin{figure}
\begin{center}
 \includegraphics[width=\columnwidth]{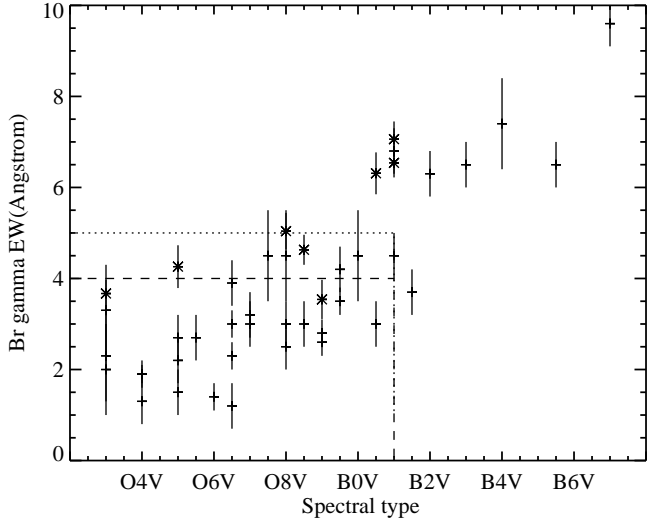}
  \caption{The EW of \brg\ plotted versus spectral type. The points
  represented by the ``+'' signs are taken from \citet{Hanson96}. The
  EWs indicated with a ``$\star$'' are from our comparison spectra in
  Fig. \ref{fig:refstars} (see also Table \ref{tab:refstars}). The
  dashed line represents the criterion of 4~\AA\ to discriminate between the
  kO9--B1 and kB2--B3 spectral classes. However, a lot of O8V--B0V
  stars have \brg\ EWs larger than 4~\AA\ and would be wrongly classified as
  kB2-B3. By shifting the criterion to 5~\AA, the O8V--B0V stars
  would be correctly classified.\label{fig:EWplot}}
\end{center}
\end{figure}

Another adaptation we applied to the classification scheme of
\citet{Hanson96} is the division between the kO9--B1 and kB2--B3
spectral classes. The difference between the kO9--B1 and kB2--B3
spectral classes is only based on the strength of the \brg\ line; for
both spectral classes the \hei\ line is of comparable strength.  The
division between those two classes is set at a \brg\ EW of 4~\AA. A
substantial number of O8V--B1V stars, however, have a \brg\ EW between
4 and 5~\AA\ (Fig. \ref{fig:EWplot}): The EW of the \brg\ line in the
O8-O9 stars \object{HD 93343} (O8V), \object{CCDM J10452-5945A}
(O8.5V) and \object{HD 326329} (O9V) is 5.0, 4.6, and 3.5
respectively.  By shifting the division to 5~\AA, the O8V-B0V stars
are classified as kO9-B1 stars.  Ideally, line-ratios are much better
suited for spectral classification. However, due to the lack of lines
in the late-O and early-B stars, the EW of \brg\ has to be used.  This
problem does not occur when discriminating kB2--B3 and kB4--B7 as in
between these two spectral classes the \hei\ line disappears.

\citet{Hanson96} do not find a significant
difference in spectral features between the dwarfs (luminosity class V) and
the giants (III).  The supergiants, however, show a different spectrum,
especially in the \brg\ and helium lines, where the stronger stellar
wind produces emission lines (or wind emission fills in the absorption lines).

\subsection{Spectral classification of embedded OB stars \label{sec:ourspectra}}

\begin{figure*}[!ht]
 \includegraphics[width=2\columnwidth]{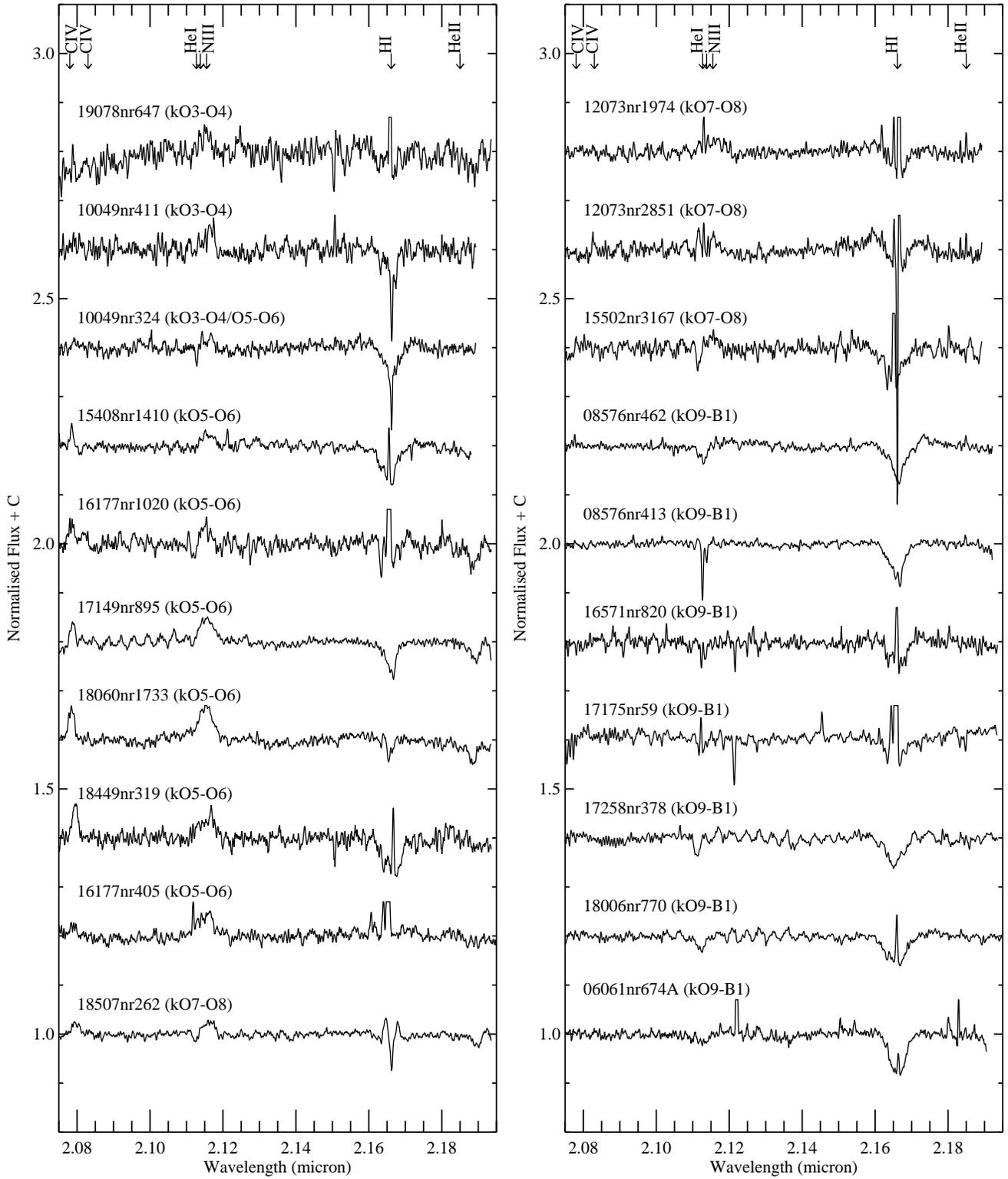}
  \caption{$K$-band spectra of young OB stars deeply embedded in
  star-forming regions. The objects are plotted according to their
  $K$-band spectral type. In the early O stars (left panel) \civ\ and
  \niii\ are important diagnostics, as well as the \heii\ line. At
  mid-O spectral type, the \hei\ line becomes an important diagnostic
  together with the \brg\ line (right panel). The emission wings
  present in the \brg\ line of 12073nr2851 are likely the result of a
  bad correction of the \brg\ line in the standard
  star. \label{fig:ostars1and2}}
\end{figure*}

\begin{figure*}[!ht]
 \includegraphics[width=2\columnwidth]{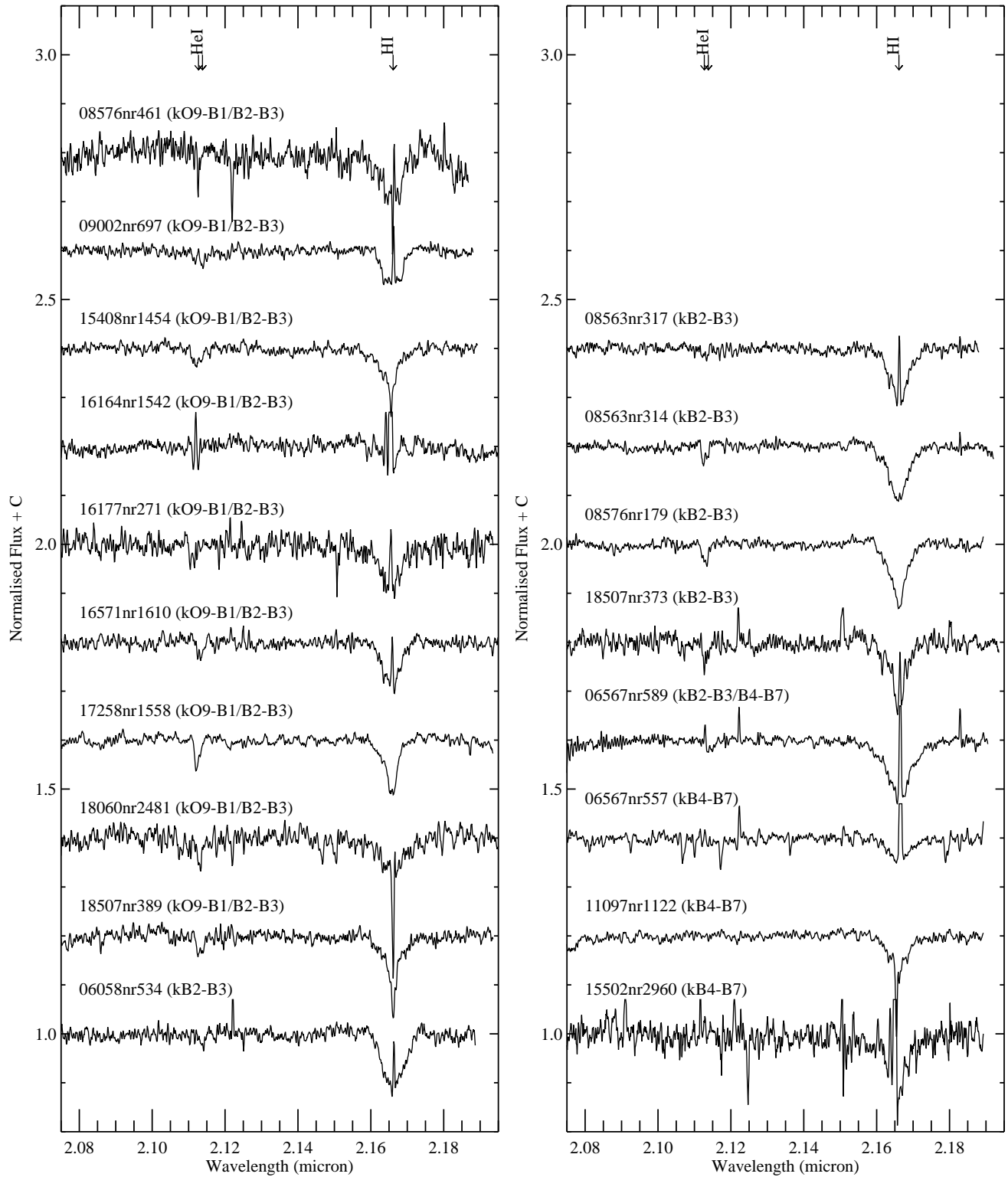}
  \caption{$K$-band spectra of young OB stars deeply embedded in
  star-forming regions, continued. In this figure the B stars are
  plotted. In early-B stars, the \hei\ line is still present (left
  panel), but in the mid-B stars the \brg\ line is the only diagnostic
  line. \label{fig:ostars3and4}}
\end{figure*}

The $K$-band spectra of the embedded OB stars are presented in
Figs. \ref{fig:ostars1and2} and \ref{fig:ostars3and4}. In the
remaining part of the paper we will name the objects after the first 5
digits (i.e. the right ascension) of the IRAS point source they are
associated with, together with a number based on our photometry
\citep[e.g. object 647 in \object{IRAS 19078+0901} we refer to as
\object{19078nr647}, cf.][]{Kaper04}. Two spectra include strong
emission lines characteristic of OB supergiants and WR stars
(Fig. \ref{fig:ostarsem}). Because of the presence of photospheric
lines, like \niii\ and \civ, these objects were not included in the
massive YSO sample \citep{Brgspec04} even though they show \brg\ in
emission.  The spectra in Figs. \ref{fig:ostars1and2} and
\ref{fig:ostars3and4} are plotted according to their $K$-band spectral
type (see Table \ref{tab:distance} and below).

\begin{table*}
\begin{center}
\begin{tabular}{l|lllll}\hline 
\multicolumn{1}{c}{Object}&\multicolumn{1}{c}{\civ}&\multicolumn{1}{c}{\niii} &\multicolumn{1}{c}{\hei}&\multicolumn{1}{c}{\ion{H}{i}} &\multicolumn{1}{c}{\heii}  \\ \hline
\noalign{\smallskip}
\object{06058nr534}     & ...             & ...            &  $0.38 \pm 0.15$  & $7.6  \pm 0.7$  &  n \\
\object{06061nr674A}    & ...             & ...            &  $0.60 \pm 0.19$  & $4.1  \pm 0.3$  &  n  \\
\object{06567nr557}     & ...             & ...            &  ...               & $3.5  \pm 0.6$  &  n  \\
\object{06567nr589}     & ...             & ...            &  $0.43 \pm 0.16$  & $10.3 \pm 0.6$  &  n   \\
\object{08563nr314}     & ...             & ...            &  $0.61 \pm 0.12$  & $7.0  \pm 0.4$  &  n  \\ 
\object{08563nr317}     & ...             & ...            &  $0.34 \pm 0.14$  & $6.5  \pm 0.4$  &  n  \\ 
\object{08576nr179}     & ...             & ...            &  $0.71 \pm 0.11$  & $6.6  \pm 0.3$  &  n   \\ 
\object{08576nr413}     & ...             & ...            &  $0.86 \pm 0.03$  & $3.7  \pm 0.2$  &  n   \\
\object{08576nr461}     & ...             & ...            &  $0.57 \pm 0.11$  & $5.2  \pm 1.0$  &  n  \\
\object{08576nr462}     & ...             & ...            &  $0.68 \pm 0.16$  & $3.1  \pm 0.3$  &  n   \\ 
\object{09002nr697}     & ...             & ...            &  $0.70 \pm 0.24$  & $4.9  \pm 0.4$  &  n \\
\object{10049nr324}     & $-0.22\pm0.12$ &$-0.67\pm0.16$ &  $0.22 \pm 0.05$  & $3.0  \pm 0.3$  &  n   \\ 
\object{10049nr411}     & ...             &$-1.8 \pm 0.5$ & ...                & $2.4  \pm 0.6$  &  n \\ 
\object{11097nr1122}    & ...             & ...            & ...                & $4.3  \pm 0.4$  &  n   \\
\object{12073nr1974}    &                &$-1.6\pm 0.4 $ & neb               & $2.5  \pm 0.3$  &  n  \\ 
\object{12073nr2851}    & ...             &$-1.8\pm 0.2$  & neb               & $2.6  \pm 0.4$  &  n  \\ 
\object{15408nr1410}    & $-0.57\pm0.08$ &$-1.20\pm0.28$ & ...                & $3.0  \pm 0.3$  &  n  \\ 
\object{15408nr1454}    & ...             & ...            &  $1.0  \pm 0.22$  & $4.8  \pm 0.5$  &  n  \\ 
\object{15502nr2960}    &  ...            & ...            & ...                & $6.1  \pm 1.0$  &  n  \\
\object{15502nr3167}    & ...             &$-0.71\pm0.35$ &  $0.64 \pm 0.14$  & $4.9  \pm 0.6$  &   n \\
\object{16164nr1542}    &  ...            & ...            &  $1.1  \pm 0.3 $  & $4.5  \pm 1.5$  & ...      \\
\object{16177nr271}     & ...             &...             & $0.83 \pm 0.35$   & $5.4  \pm 0.9$  & ...   \\ 
\object{16177nr405}     & $-0.55\pm0.22$ &$-2.3\pm0.4$   & neb               & ...       & ...    \\ 
\object{16177nr1020}    & $-1.0\pm0.25$  &$-1.7\pm0.4$   & ...                & $2.7  \pm 0.5$  & $1.45 \pm 0.35$\\ 
\object{16571nr820}     & ...             &...             & $0.57 \pm 0.1$    & $2.4  \pm 0.4$  & ...   \\ 
\object{16571nr1610}    & ...             & ...            & $0.66 \pm 0.14$   & $5.2  \pm 0.4$  & ...      \\ 
\object{17149nr895}     & $-0.87\pm0.09$ & $-2.5\pm0.3$  & ...                & $2.0  \pm 0.2$  & $0.95 \pm 0.16$ \\
\object{17175nr59}      & ...             &               & $0.86 \pm 0.14$   & $4.4  \pm 0.5$  & ...    \\
\object{17258nr378}     &  ...            & ...            & $0.73 \pm 0.14$   & $3.0  \pm 0.4$  & ...      \\
\object{17258nr1558}    &  ...            & ...            & $0.95 \pm 0.10$   & $4.9  \pm 0.3$  & ...   \\
\object{17423nr3102}    & $-1.0 \pm 0.16$& $-6.1\pm 0.5$ & blend  & $25.4 \pm 0.2$  & $0.6 \pm 0.16$ \\
\object{18006nr770}     & ...             & ...            & $0.75 \pm 0.36$   & $3.6  \pm 0.5$  & ...      \\ 
\object{18060nr1733}    & $-1.2 \pm 0.24$& $-3.5\pm0.4$  & ...                & $0.76 \pm 0.2$  & $0.76 \pm 0.16$  \\ 
\object{18060nr2481}    &                &               & $0.75 \pm 0.22$   & $4.9  \pm 1.0$  & ...      \\
\object{18449nr319}     & $-1.4\pm0.2$   & $-2.7 \pm0.7$ & ...                & $3.9  \pm 0.6$  & ...  \\ 
\object{18449nr335}     &  ...            & $-10.6\pm 0.5$& blend  & $-44.2\pm 1.0$  & $-3.2 \pm 0.2$ \\
\object{18507nr262}     & $-0.7\pm0.15$  & $-1.1 \pm0.2$ & $0.12\pm0.06$     & ??              & $0.5 \pm 0.15$   \\
\object{18507nr373}     & ...             &...             & $0.79\pm0.27$     & $6.6  \pm 0.5$  & ...      \\ 
\object{18507nr389}     & ...             &...             & $0.79\pm0.26$     & $5.3  \pm 0.3$  & ...   \\ 
\object{19078nr647}     & ...             & $-1.9 \pm 0.6$& ...                & $4.0 \pm 1.2$   & $1.3 \pm 0.5$ \\ \hline  
\end{tabular}
\caption{EW measurements of the spectral lines detected in the
$K$-band spectra of young OB stars. Before measuring the EW, the
nebular \brg\ and \hei\ lines have been subtracted out, if present.
The \heii\ line is not covered in the wavelength setting used in
visitor mode (indicated with \emph{n} in column 6). Emission lines
have a negative EW. The error on the EW takes into account the
signal-to-noise ratio and the peak over continuum ratio of the
line. \label{tab:EW}}
\end{center}
\end{table*}

The EW measurements are presented in Table \ref{tab:EW} in order of
right ascension of the IRAS source. In some spectra, the \hei\ and/or
the \brg\ lines are suffering from nebular contamination. In most of
these, the lines have been subtracted out before measuring the \hei\
or \brg\ EW. In a few cases the nebular
\hei\ emission was so strong that the underlying absorption could not
be detected any longer (noted in Table \ref{tab:EW} as
``neb''). This problem was less severe in case of \brg\ as the
photospheric absorption line is much broader than the nebular emission
line. Also the \hei\ nebular lines at 2.161 and 2.165 \micron\ which
are sometimes present bluewards of \brg\ do not prevent the detection
of the broad \brg\ absorption.  The error on the EW is determined
taking into account the signal-to-noise ratio of the spectrum (see column 5,
Table \ref{tab:obslog}) and the peak over continuum ratio of the line.

The $K$-band spectra are classified based on the method outlined in
Sect. \ref{sec:classification}. The resulting spectral class and the
corresponding range in optical spectral type is given in column 2 and
3 of Table \ref{tab:distance}. As described in
Sect. \ref{sec:classification}, the difference between kO9--B1 and
kB2--B3 is only the strength of the \brg\ line. An EW of 5 \AA\ is
used to discriminate between these two spectral types. The spectra
where the EW measurement of \brg\ is equal (within the errors) to 5
\AA\ are classified as kO9--B1/kB2--B3.

In \object{16177nr405} (Fig. \ref{fig:ostars1and2}, left panel), the
\brg\ absorption line is not detected. The spectral type derived for
this object is based on the presence of the \civ\ and \niii\
lines. The \hei\ line, if present, could not be detected because of
strong nebular emission. For this spectral type (kO5-O6), \heii\
should also be present in the spectrum. The \heii\ line, however, is
hardly detected, a little hint of absorption may be present.  Also in
\object{18449nr319} (Fig. \ref{fig:ostars1and2}, left panel) the
\heii\ line is not detected.  Data taken at a second epoch
\citep{ApaiThesis04} indicate that the \heii\ line in the spectrum of
\object{18449nr319} is present. This suggests that an instrumental
effect may be responsible for these non-detections. In
\object{16177nr405}, however, the \heii\ line is absent in both
epochs.  The \brg\ line in \object{18507nr262}
(Fig. \ref{fig:ostars1and2}, left panel) shows a remarkable
profile. The absorption in the center of the line is due to
over-subtraction of the spatially extended nebular \brg\ emission. The
emission wings, however, are real, suggesting that this \brg\ line is
in emission. The full width at zero intensity (FWZI) is $\sim
800$~\kms.

\subsection{Emission line objects \label{sec:emission}}
\begin{figure}
\begin{center}
 \includegraphics[width=\columnwidth]{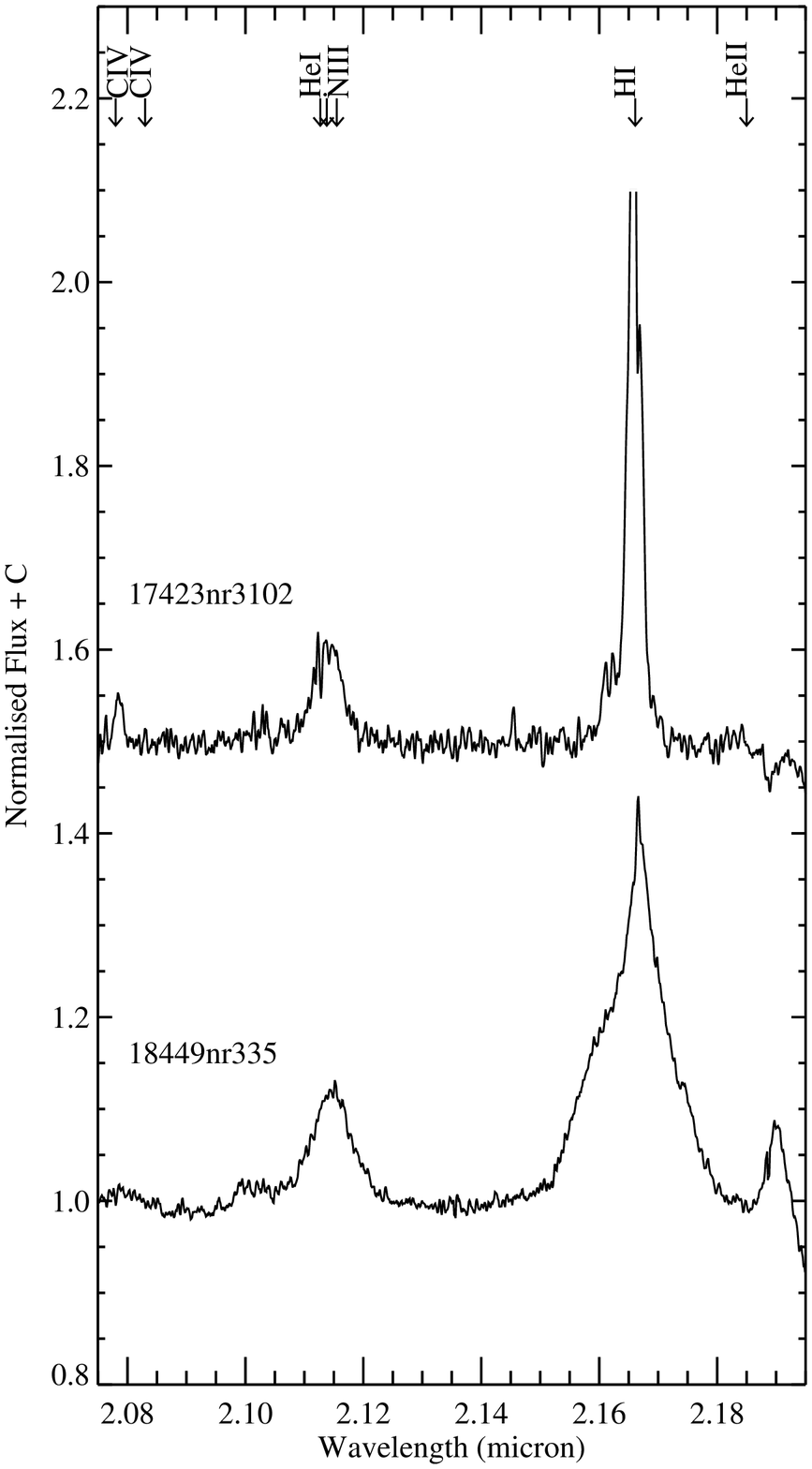}
   \caption{$K$-band spectra of two early-type stars showing emission
   lines. The spectrum of \object{17423nr3102} shows the photospheric
   features of a kO5-O6b star, the \brg\ emission line likely indicates a
   strong stellar wind. The bottom spectrum (\object{18449nr335}) has
   the characteristics of a WN star. \citet{Blum99} classify the spectrum as WN7.}
   \label{fig:ostarsem}
\end{center}
\end{figure}

The spectra of \object{17423nr3102} and \object{18449nr335}
clearly show \brg\ in emission (Fig. \ref{fig:ostarsem}). These stars
do not share the other spectral properties of massive YSOs, which do
show \brg\ in emission, but no \civ\ and \niii\ emission nor \heii\
absorption.  In the $K$-band spectrum of
\object{17423nr3102} \civ\ and \niii\ are present, as well as \heii\
absorption. These lines are formed in the photosphere of the
star. \citet{Cotera99} classify this star as a Be star, based on a
low-resolution $K$-band spectrum. This classification can be excluded by our 
high-resolution spectrum. The $K$-band spectra of Be stars do not show
\civ\ and \niii\ emission lines \citep{Clark00}. Applying the
classification scheme of \citet{Hanson96} the source would be
classified as kO5-O6b, an O5-O6.5 supergiant.

The $K$-band spectrum of \object{18449nr335} displays 
very  broad \brg\ emission profiles with a triangular shape,
characteristic of Wolf-Rayet stars.
This star is located in the center of \object{W43} and is
classified by \citet{Blum99} as a WN7 star.

\begin{table*}
\begin{tabular}{lllllllll}\hline 
\multicolumn{1}{c}{Object}&\multicolumn{1}{c}{$K$-band}&\multicolumn{1}{c}{MK} &\multicolumn{1}{c}{Kin. dist } &\multicolumn{1}{c}{Dist.} & \multicolumn{1}{c}{$K$}&\multicolumn{1}{c}{$J-K$} & \multicolumn{1}{c}{$A_\mathrm{V}$} &\multicolumn{1}{c}{Sp. Ph. dist.}   \\
&\multicolumn{1}{c}{sp. type}&\multicolumn{1}{c}{sp. type} &\multicolumn{1}{c}{(kpc)} &\multicolumn{1}{c}{(kpc)}&\multicolumn{1}{c}{} &\multicolumn{1}{c}{} & &\multicolumn{1}{c}{(kpc)}    \\ \hline \hline
\noalign{\smallskip}
\object{06058nr534}   &  kB2--B3                & B1V - B2.5V    &  1.0        &  2.2          &10.4$\pm$ 0.02  &  1.3$\pm$0.04    & 7.3      &  1.0 --  1.5 \\
\object{06061nr674A}  &  kO9--B1                & O8V - B1V      &  --         &  2.2          &11.8$\pm$0.04   &  2.0$\pm$0.09    & 11.4     &  2.3 --  5.1  \\
\object{06567nr557}   &  kB4--B7                & B3V - B7V      &  2.6        &  2.3          &11.6$\pm$0.04   &  2.6$\pm$0.10    & 14.0     &  0.8 --  1.1  \\
\object{06567nr589}   &  kB2--B3/kB4--B7        & B1V - B7V      &  2.6        &  2.3          &10.4$\pm$0.02   &  3.2$\pm$0.07    & 17.3     &  0.4 --  0.8   \\
\object{08563nr314}   &  kB2--B3                & B1V - B2.5V    &  2.1        &  2.0          &8.9 $\pm$0.01   &  1.4$\pm$0.02    & 7.8      &  0.5 --  0.7  \\
\object{08563nr317}   &  kB2--B3                & B1V - B2.5V    &  2.1        &  2.0          &10.0 $\pm$0.02  &  2.3$\pm$0.04    & 12.6     &  0.6 --  0.9  \\
\object{08576nr179}   &  kB2--B3                & B1V - B2.5V    &  2.2        &  0.7          &9.6$\pm$0.01    &  1.9$\pm$0.03    & 10.5     &  0.6 --  0.8  \\
\object{08576nr413}   &  kO9--B1                & O8V - B1V      &  2.2        &  0.7          &7.5$\pm$0.01    &  1.8$\pm$0.01    & 10.4     &  0.3 --  0.8  \\
\object{08576nr461}   &  kO9--B1/kB2--B3        & O8V - B2.5V    &  2.2        &  0.7          &12.7$\pm$0.06   &  1.7$\pm$0.11    &  9.8     &  2.6 --  8.5  \\
\object{08576nr462}   &  kO9--B1                & O8V - B2.5V    &  2.2        &  0.7          &7.0$\pm$0.01    &  1.8$\pm$0.01    & 10.4     &  0.2 --  0.6  \\
\object{09002nr697}   &  kO9--B1/kB2--B3        & O8V - B2.5V    &  1.9        &  2.0         &10.7$\pm$0.03   &  5.4$\pm$0.22    & 29.0      &  0.4 --  1.2 \\
\object{10049nr324}   &  kO3--O4/kO5--O6        & O3V - O6.5V    &  7.1        &  7.1           &10.5$\pm$0.02   &  2.8$\pm$0.06    & 15.7    &  2.7 --  4.4 \\
\object{10049nr411}   &  kO3--O4                & O3V - O4V      &  7.1        &  7.1          &11.7$\pm$0.04   &  2.9$\pm$0.11    & 16.2     &  6.5 --  7.4  \\
\object{11097nr1122}  &  kB4--B7                & B3V - B7V      &  3.1        &  2.8          &10.5$\pm$0.02   &  2.4$\pm$0.05    & 13.0     &  0.5 --  0.7  \\
\object{12073nr1974}  &  kO7--O8                & O5V - O6.5V    &  11.6       &  11.6         &10.7$\pm$0.03   &  2.3$\pm$0.05    & 13.0     &  2.8 --  3.2  \\
\object{12073nr2851}  &  kO7--O8                & O7V - O8V      &  11.6       &  11.6         &10.7$\pm$0.03   &  2.5$\pm$0.06    & 14.1     &  2.7 --  3.0 \\ 
\object{15408nr1410}  &  kO5--O6                & O5V - O6.5V 	 &  2.6/11.6   &  2.6            &8.6$\pm$0.01    &  2.2$\pm$0.02    & 12.0   &  1.9 --  2.4  \\
\object{15408nr1454}  &  kO9--B1/kB2--B3        & O8V - B2.5V    &  2.6/11.6   &  2.6           &9.3$\pm$0.01    &  2.1$\pm$0.03    & 12.5    &  0.3 --  1.1 \\
\object{15502nr2960}  &  kB4--B7                & B3V - B7V      &  5.8/8.7    &  5.8            &13.0$\pm$0.08   &  2.6$\pm$0.19    & 14.0   &  1.5 --  2.1 \\
\object{15502nr3167}  &  kO7--O8                & O7V - O8V      &  5.8/8.7    &  5.8            &11.8$\pm$0.04   &  1.6$\pm$0.19    &  9.3   &  5.8 --  6.6 \\
\object{16164nr1542}  &  kO9--B1/kB2--B3        & O8V - B2.5V    &  3.7/11.4   &  3.6           &11.7$\pm$0.04   &  2.0$\pm$0.08    & 11.4    &  1.5 --  4.9 \\
\object{16177nr271}   &  kO9--B1/kB2--B3        & O8V - B2.5V    &  3.4/11.8   &  3.6           &11.8$\pm$0.04   &  3.4$\pm$0.14    & 18.8    &  1.1 --  3.4 \\
\object{16177nr405}   &  kO5--O6                & O5V - O6.5V    &  3.4/11.8   &  3.6           &10.8$\pm$0.03   &  $\ge$6.7 	    & $>$35.4 &  $>$1.0        \\
\object{16177nr1020}  &  kO5--O6                & O5V - O6.5V    &  3.4/11.8   &  3.6           &11.9$\pm$0.04   &  4.7$\pm$0.27    & 25.5    &  3.0 --  3.7 \\
\object{16571nr820}   &  kO9--B1                & O8V - B1V      &  1.9/14.5   &  1.0           & 9.3$\pm$0.01   &  2.6$\pm$0.03    & 14.6    &  0.6 --  1.4 \\    
\object{16571nr1610}  &  kO9--B1/kB2--B3        & O8V - B2.5V    &  1.9/14.5   &  1.0           &11.0$\pm$0.03   &  2.5$\pm$0.07    & 14.1    &  1.0 --  3.1 \\
\object{17149nr895}   &  kO5--O6                & O5V - O6.5V    &  2.0/14.6   & 2.0           &8.3 $\pm$0.01   &  0.6$\pm$0.01    & 3.9      &  1.9 --  2.3  \\
\object{17175nr57}    &  kO9--B1                & O8V - B1V      &  2.0/14.8   &  2.0         &11.5$\pm$0.03   &  6.3$\pm$0.5     & $>$33.4   &  0.6 --  4.1 \\
\object{17258nr378}   &  kO9--B1                & O8V - B1V      &  2.3/14.6   & 2.3          &10.2$\pm$0.02   &  3.4$\pm$0.06    & 18.8      &  0.7 --  1.6  \\
\object{17258nr1558}  &  kO9--B1/kB2--B3        & O8V - B2.5V    &  2.3/14.6   & 2.3           &10.7$\pm$0.02   &  1.9$\pm$0.04    & 10.9     &  1.0 --  3.2  \\
\object{17423nr3102}  &  kO5--O6b               & O5I - O6I      & $^\dagger$  & 8.5           &9.9$\pm$0.02    &  5.3$\pm$0.13    & 28.5     &  2.6 --  2.6   \\
\object{18006nr770}   &  kO9--B1                & O8V - B1V      &  1.2        & 1.9           &7.3 $\pm$0.01   &  0.9$\pm$0.01    & 5.5      &  0.4 --  0.9  \\
\object{18060nr1733}  &  kO5--O6                & O5V - O6V      &  2.1/14.6   &  2.1          &9.0$\pm$ 0.01   &  2.9$\pm$0.03    & 16.2     &  1.4 --  1.6   \\  
\object{18060nr2481}  &  kO9--B1/kB2-B3         & O8V - B2.5V    &  2.1/14.6   &  2.1          &11.1$\pm$0.03   &  3.5$\pm$0.11    & 19.3     &  0.7 --  2.4    \\
\object{18449nr319}   &  kO5--O6                & O5V - O6V      &  5.9/8.7    & 4.3           &10.8$\pm$0.03   &  2.4$\pm$0.06    & 13.5     &  3.8 --  4.3   \\
\object{18449nr335}   &  WN7                    & WN7            &  5.9/8.7    & 4.3           &8.5$\pm$0.01    &  7.0$\pm$0.17    & --       &                \\
\object{18507nr262}   &  kO5--O6                & O5V - O8V      &  3.7/10.3   & 3.7           &9.4$\pm$0.01    &  4.3 $\pm$0.07   & 23.4     &  0.9 --  1.3     \\ 
\object{18507nr373}   &  kB2--B3                & B1V - B2.5V    &  3.7/10.3   & 3.7           &12.3$\pm$0.05   &  5.4$\pm$0.49    & 28.6     &  0.8 --  1.1  \\
\object{18507nr389}   &  kO9--B1/kB2--B3        & O8V - B2.5V    &  3.7/10.3   & 3.7           &11.3$\pm$0.03   &  3.8 $\pm$0.14   & 20.9     &  0.7 --  2.4   \\
\object{19078nr647}   &  kO3--O4                & O3V - O4V      &  0.1/12.3   & 11.4          &13.2$\pm$0.07   &  4.4 $\pm$0.5    & 23.9     &  8.4 --  9.6   \\\hline
\end{tabular}
\caption{Classification of the $K$-band spectra based on  EW
  measurements (Table \ref{tab:EW}) and classification scheme
  (Table \ref{tab:scheme}) developed by \citet{Hanson96}. Column 2:
  the $K$-band spectral type, column 3: the corresponding optical
  spectral type.  Column 4: kinematic distance derived from the
  radial velocity measurements of \citet{Bronfman96} and the rotation
  model of the galaxy of \citet{Brand93}. Column 5: adopted distance
  \citep{Kaper04}.  Column 6: $K$-band magnitude; column 7: $J-K$
  ; column 8: the derived visual extinction
  ($A_\mathrm{V}$). Column 9, the distance based on this spectral
  classification, assuming that the stars are ZAMS stars
  \citep{Hanson97}.  $^\dagger$: This object is located towards the
  galactic center. \label{tab:distance} }



\end{table*}


\subsection{Distance determinations}\label{sec:distance}

The kinematic distance of the UCHII regions is listed in column 4 of
Table \ref{tab:distance}.  These distances are derived from the radial
velocity measurements of the CS(2--1) line \citep{Bronfman96}. A model
for the galactic rotation developed by \citet{Brand93} is used to
convert the radial velocity into a distance estimate. For sources
located inside the solar-circle, a near and a far distance is
given. In column 5 of Table \ref{tab:distance} the distance adopted in
\citet{Kaper04} is given.  These estimates are obtained from the
literature and based on, e.g., star counts, radio recombination lines
and optical photometry.

These two distance estimates can be compared with the distance
derived from the $K$-band spectral type and photometry
(spectro-photometric distance or spectroscopic paralax). By assuming that the stars are all
located on the ZAMS, the intrinsic absolute magnitude can be
derived. The absolute $K$-band magnitudes for ZAMS stars are taken from
\citet{Hanson97}, who follow \citet{Massey89} for
spectral types between O3V and B1V and \citet{Kenyon95} for the B stars.
The $J-K$ colour is almost constant for OB stars \citep[$J-K =
-0.2$,][]{Koornneef83}, allowing a good determination of the
extinction towards the OB star.

By comparing the different distance estimates, for
more than half of the stars ($\sim$ 60\%) the spectro-photometric
distance is comparable to the kinematic distance. In a few cases
(e.g. \object{IRAS 16177-5018} and \object{IRAS 17258-3637}) the
distance ambiguity can be resolved.  For other objects, the literature
distance based on optical photometry shows a better match with the
spectro-photometric distance (objects in \object{Vela
C}). \citet{Murphy91} show that \object{IRAS 08563-4711} and
\object{IRAS 09002-4732} are both located near the edge of the more
distant \object{Vela B} cloud (2 kpc). The spectro-photometric
distance, however, suggests that \object{IRAS 08563-4711} and
\object{IRAS 09002-4732} are located in \object{Vela C} (0.7 kpc).

For the other  objects, the
spectro-photometric distance and the kinematic distance are not consistent
with each other. \citet{Brand93} show that the residual radial velocities can be large
(upto 40 \kms) in certain directions, and will introduce a large uncertainty
in the determination of the kinematic distance.


Also the  distances derived for stars in the same
embedded cluster can be different. In four  regions, the
spectro-photometric distance of one of the stars is not consistent
with the distance derived for the other stars. For example, in 
\object{IRAS 10049-5657}, the spectral type of the two stars
(\object{10049nr324} and \object{10049nr411}) is similar, but the
$K$-band magnitude differs by 1 magnitude. The two stars have a
similar $J-K$ suggesting that they are confronted with the same amount of
extinction, which makes it unlikely that one of the stars is a
foreground star (the spectro-photometric distance towards the faintest
star, \object{10049nr411}, is similar to the kinematic distance). The
magnitude difference could be explained if \object{10049nr324}
is an equal-mass binary (2 O3-O4 stars) which would result in an increase
by 0.7 magnitude.

Also in \object{08576-4334} we derive for one of the stars a different
spectro-photometric distance. In this cluster we took spectra of 4
late-O, early B stars. Three stars have a spectro-photometric distance
consistent with the distance to the \object{Vela Molecular
Ridge}. Like in \object{IRAS 10049-5657}, all the objects are
suffering from the same amount of extinction (column 8, Table
\ref{tab:distance}). The magnitude of \object{08576nr461} is $3-4$
magnitudes too faint. Although the spectrum has a low S/N, the \hei\
line is detected, which determines the spectral type to be
kO9-B1/kB2-B3. Even if the \hei\ line were not present and the
star would be classified as kB4-B7, the distance does not match. It
likely is a background star; in this direction, one looks along a
spiral arm resulting in the projection of several star-forming regions.
Also for \object{15408nr1454} and \object{15502nr2960} the derived distances are
not consistent with the distances for the other stars in the same cluster.

\section{$K$-band spectra of near-infrared counterparts of
  UCHII regions \label{sec:uchiicp}}

The comparison done in \citet{Kaper04} between the near-infrared
images and the UCHII radio observations \citep[ and Kurtz,
priv. comm.]{WoodRadio89,Kurtz94,Walsh98} shows that 21 UCHII radio
sources have a near-infrared counterpart.  For 7 of these regions the
near-infrared counterpart is a compact cluster of stars, and the
radio emission is extended. The linear size derived for some of these
objects suggests that they are not in the UCHII stage anymore.
For instance, by adopting the distance determination of 7.3 kpc
to \object{IRAS 06412-0105}, the extent of the radio emission is
$\sim 1$ pc. For \object{IRAS 12073-6233},
which is at 11.6 kpc, a size of $6 \times 4$ pc is found (i.e. much
larger than 0.1 pc).

\begin{figure*}
 \includegraphics[width=2.0\columnwidth]{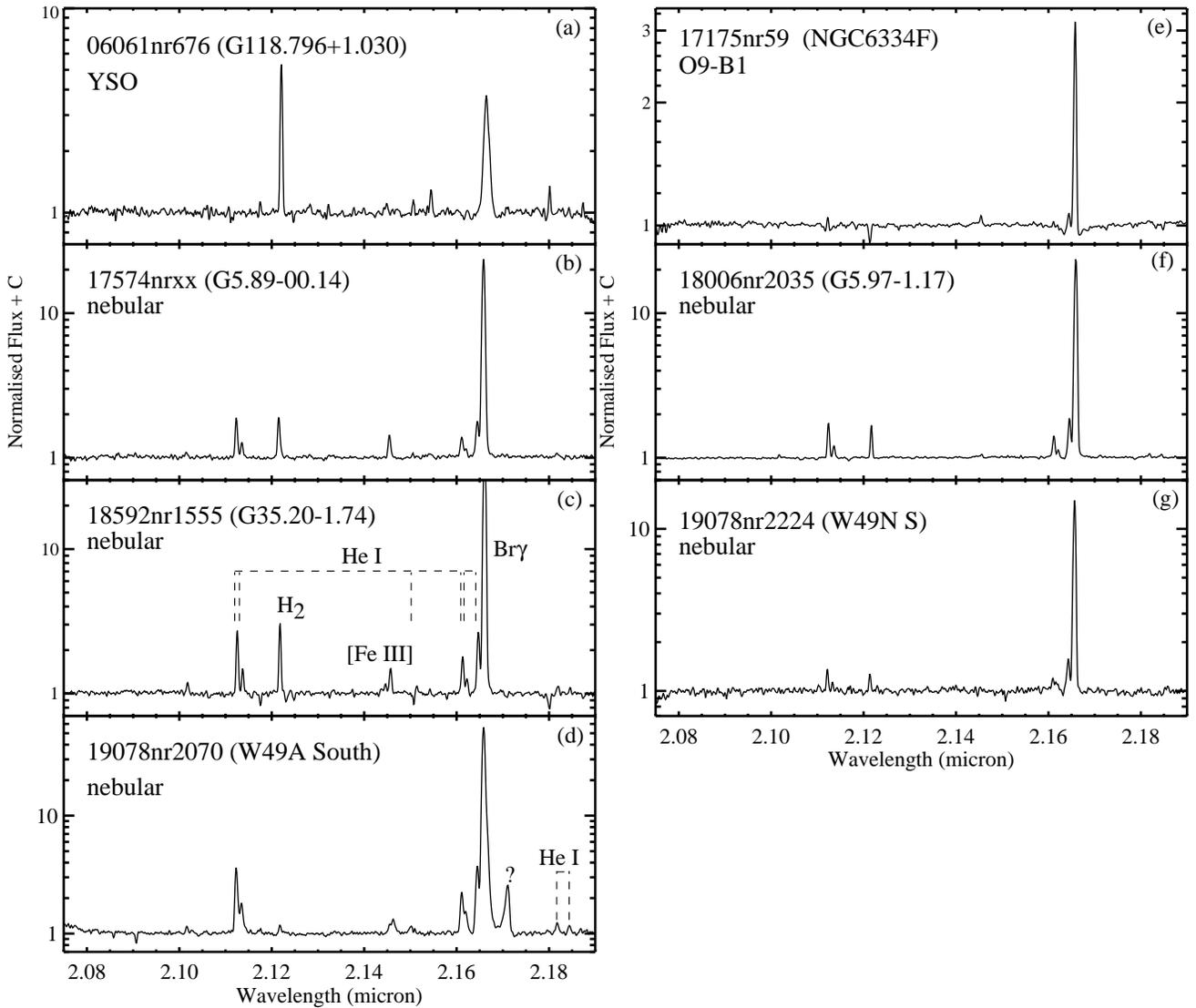}
   \caption{$K$-band spectra of the near-infrared counterpart of
   ultra-compact radio source. In the upper-left corner of each panel
   the name of the source is given with the name of the UCHII radio
   source in parentheses. The flux scale is logarithmic.  The line
   identifications are given in panels (c) and (e). The line at 2.171
   \micron\ in panel (e) could not be
   identified. \label{fig:uchiicp}}
\end{figure*}

Towards the other UCHII radio sources, no cluster is detected, but the
near-infrared counterparts are point sources. For 7 of these objects
we have taken $K$-band spectra which are shown in
Fig. \ref{fig:uchiicp}.  All but one spectra are dominated by nebular
emission. Strong, narrow, \brg\ and \hei\ emission are seen in the
spectra.  A more careful inspection of these seven spectra reveals that
in one case (17175nr59) weak \brg\ photospheric absorption is present.
This star is classified as a kO9-B1. Another object (06061nr676)
displays a broad and spectrally resolved \brg\ line, and has the
characteristics of a massive Young Stellar Object; this object is
discussed in detail in \citet{Brgspec04}. For comparison we include
these objects in Fig. \ref{fig:uchiicp}.

Comparison of the magnitudes and colours of the three types of objects
(OB stars, nebular spectra and massive YSO candidates) found in our
spectroscopy survey shows that the near-infrared counterparts of UCHII
regions have on average the faintest $K$ magnitude and the reddest
$J-K$ colour. This suggests that these sources suffer from more
extinction than the OB stars and massive YSOs do. The near-infrared
continuum of the counterparts of the UCHII regions is likely not
dominated by the stellar continuum as no photospheric features are
detected. The slope of the continuum is likely much redder than that
of the OB stars; the continuum will be caused by free-free emission of
the UCHII region in combination with hot dust heated by the ionising
star in the UCHII region or scattered light.

Since the UCHII regions are often found in areas with extended radio
and associated near-infrared line emission from a larger,
(evolved) \HII\ region, the question arises what the origin is of the
nebular emission lines detected towards the UCHII regions. Inspection
of the spatial distribution of the nebular emission of our 7 sources
indicates that in almost all cases the line emission is peaked at the
position of the radio source. The \hei\ emission, if present, also
peaks  at the radio source. 

In addition, in one case we can compare the width of the radio
recombination lines with that seen in our spectra. For \object{W49A
South} our spectrum indicates a width of 90 \kms, in agreement with
the width seen at radio frequencies \citep{Depree97}. We conclude that
there is good reason to assume that at least in some UCHII regions the
near-infrared emission lines are caused by the same gas producing the
radio emission.

Therefore, the nebular lines originating from the UCHII region can be
used to estimate the spectral type of the underlying ionising
source. The presence of \hei\ is indicative of stars with a spectral
type earlier than O9V \citep{Hanson02}. This is confirmed by the weak
\hei\ line present in the spectrum of the O9-B1V star
(\object{17175nr59}). This means that, according to their strong \hei\
emission, most of the stars embedded inside the UCHII regions are of
early O spectral type.  This preference to detect the hottest O stars
is not surprising, since these objects will produce the brightest
UCHII regions and will also produce strong nebular line emission in
the near-infrared. In contrast, late O or B stars will in general not
be easily  detected as UCHII regions. Their immediate environment,
however, will contain remnants of the star formation process for a
much longer period of time than that of the hottest O stars. Indeed,
the YSO candidates are often found to be of B spectral type and are
generally not UCHII regions.

\section{Ionising properties of the embedded clusters}\label{sec:regions}

One of the results of the near-infrared imaging survey of
\citet{Kaper04} is that the UCHII regions are not isolated, but are
located inside more extended regions of (massive) star formation. In
almost all these regions embedded clusters are detected. 
 The properties of these clusters are discussed in detail in
\citet{Kaper04}. The majority of the spectra presented in this paper do not
correspond to the near-infrared counterpart of the UCHII region
detected in the radio, but to massive stars located inside the cluster
associated with the UCHII region.

For the determination of the ionising power of these clusters, the
UCHII radio observations are not suitable. Only small spatial scales
are resolved by these interferometric radio surveys.  The radio flux of the more
extended \HII\ region is not picked up \citep[see e.g.][]{Kurtz99}.
Therefore, we use the radio measurements presented by
\citet{Caswell87}. These are single dish observations with a typical
spatial resolution of 1\arcmin, which makes the observations more
compatible with the IRAS observations. The near-infrared images
presented in \citet{Kaper04} cover an area of $5\times5\arcmin$, an
area comparable on the sky to the radio and infrared measurements.

\begin{table*}
\begin{tabular}{llllp{9.3cm}} \hline\hline
\multicolumn{1}{c}{IRAS}
&\multicolumn{1}{c}{Radio}&\multicolumn{1}{c}{IRAS} & \multicolumn{1}{c}{$K$-band} &\multicolumn{1}{c}{Ionising source} \\
\multicolumn{1}{c}{name} &\multicolumn{1}{c}{Sp Type}&\multicolumn{1}{c}{Sp Type} & \multicolumn{1}{c}{Sp. Type} &\\ \hline
\noalign{\smallskip}
\object{06058+2138} &            & B1.5V/B2V     & B1V-B2.5V   &\object{06058nr534} is the main ionising source. Massive YSOs \object{06058nr221} and \object{06058nr227} have mid - late B spectral type.  \\
\object{06061+2151} & 		 & B2V	         & O8V-B1V     & \object{06061nr674A} is the main ionising source. Spectral type is not consistent with class I nature \citep{Anandarao04}.  \\
\object{06567-0355} &  B1V(W86)	 & B1.5-B2       & B1V-B7V     & Cluster has ultra compact radio counterpart. \\
\object{08563-4711} &  O8V       & B1V           & B1V-B2.5    & two stars are ionising this region, spectral type estimates of \citet{Liseau92} (B5V and A5V) are not consistent. \\
\object{08576-4334} &  O9V	 & B2V	         & O8V-B1V     & \object{08576nr413} and \object{08576nr462} (O8V-B1V) are located in the center, massive YSOs \object{08576nr292} and \object{08576nr408} in the outskirts of the cluster. The stars in the center are likely the main ionising sources.\\
\object{09002-4732} &  O8.5V	 & O8.5V	 & O8V-B2.5V   & \citet{ApaiThesis04} conclude that this region is dominated by one ionising source coinciding with the UCHII region. \object{09002nr697} is located in the outskirts of the nebula.  \\
\object{10049-5657} &  $<$O3V	 & $<$O3V	 & O3V-O4V     & Giant
\HII\ region$^\dagger$. In the CMD of this field \citep{Kaper04} a reddened main sequence is clearly visible.  The spectral type determination of two  O3V-O4V stars (\object{10049nr411} and \object{10049nr324}) shows that this cluster must be very  massive. \\
\object{11097-6102} &  $<$O3V	 & O4V	         & B3V-B7V     & Giant \HII\ region$^\dagger$ (NGC 3576) , ionising sources are still heavily embedded and not visible in the near-infrared \citep{Figueredo02}. \\
\object{12073-6233} &  $<$O3V	 & $<$O3V	 & O7V-O8V     & Giant \HII\ region$^\dagger$. \object{12073nr1974} and \object{12073nr2851} are located in center of cluster. \\
\object{15408-5356} &  O4.5V	 & O7.5V	 & O5V-O6.5V   & \object{15408nr1410} likely the ionising source, in center of cluster.  \\
\object{15502-5302} &  O4V	 &  $<$O3V	 & O7V-O8V     & \object{15502nr3167} not the ionising source, the stars in center of cluster show nebular spectra, apparently too embedded to be observed in the near-infrared.\\
\object{16164-5046} &  O4V	 & O4V	         & O8V-B2.5V   & \object{16164nr1542} is not the ionising source. \object{16164nr3636}(YSO) might be of mid-O spectral type. \\
\object{16177-5018} &  O3V	 & O4V	         & O5V-O6.5V   & \object{16177nr1020} and \object{16177nr271} ionising sources.  \\
\object{16571-4029} &  O9V	 & B1V	         & O8V-B1V     & \object{16571nr820} and \object{16571nr1610} ionising sources. \\
\object{17149-3916} &  O8V	 & B0V	         & O5V-O6.5V   & $K$-band spectral type not consistent with other determinations. \\
\object{17175-3544} &  $<$O3V    & O8V           & O8V-B1V     & $K$-band spectral type consistent with infrared spectral type.\\
\object{17258-3637} &  O6V	 & O7.5V	 & O8V-B1V     & Main ionising source still embedded, UCHII region has no near-infrared counterpart.  \\
\object{17423-2855} & 		 & 	         &             & Located in Galactic Center \citep{Cotera99}. \\
\object{18006-2422} &   	 & O9.5V	 & O8V-B1V     & \object{18006nr770} is \object{Her 36}, ionising source of \object{M8}. \\
\object{18060-2005} &  		 & O9.5V	 & O5V-O6V     & $K$-band spectral type not consistent with infrared spectral determination.   \\
\object{18449-0158} &  		 & B0.5V	 & O5V-O6V     & \object{W43}, giant \HII\ region$^\dagger$ \citep{Blum99}. \object{18449nr319} is one of the ionising sources. \\
\object{18507+0110} &  		 & O3V	         & O5V-O6V     & Not related to UCHII region. Western part of nebula likely ionised by \object{18507nr262}\\
\object{19078+0901} &  $<$O3V    & $<$O3V        & O3V-O4V     & \object{W49A} is  one of the most luminous giant \HII\ regions$^\dagger$ in our galaxy \citep{Conti02}. \object{19078nr647} is object 1 in \citet{Conti02} and one of the ionising sources.\\ \hline
\end{tabular}	
\caption{Spectral type derived using radio and infrared diagnostics
  and compared with the OB stars found in the region by means of
  $K$-band spectroscopy. Column 1: Object; column 2: Radio spectral
  type of the extended \HII\ region based on
  \citet{Caswell87}; column 3: Infrared spectral type based on the IRAS
  flux. The IRAS luminosity is calculated using the formula of Sect. 7.6 in 
  \citet{Cox00}. The spectral types are derived using the stellar
  parameters of \citet{Smith02} for the O3 -- B1.5 stars and
  \citet{LandoltBornstein} for the B stars; Column 4: The earliest
  $K$-band spectral type of the OB star located in this region.  $\dagger$: A giant \HII\ region is defined by the fact that it emits more than 10$^{50}$ Lyman continuum photons $\rm{s}^{-1}$ \citep[e.g.][]{Blum99}.\label{tab:sptype}}

\end{table*}

  In column 2 and 3 of Table \ref{tab:sptype} the spectral types are
given based on the radio and infrared luminosity. The determination of
the radio and infrared spectral type depends on several assumptions
which do not have to be valid. E.g., one assumes that only one star is
responsible for the radio and infrared flux. The radio and infrared
spectral types have to be considered as upper limits in this case. In
the regions where the spectral type is estimated to be earlier than
O3V (e.g. IRAS 10049-5657), obviously a cluster is required to ionise
the \HII\ region and heat the surrounding dust. Based on the amount of
Lyman continuum photons ($> 10^{50}$ photons $s^{-1}$) some of these
regions are classified as giant \HII\ regions \citep[e.g.][]{Blum99}.

 Another assumption used to determine the
(radio) spectral type is that the \HII\ region is ionisation bounded (no
FUV photons are leaking) and dust-free (no FUV photons are absorbed by
dust).  For the determination of the infrared spectral type the assumption
is that all the radiation from the star is reprocessed and emitted at
infrared wavelengths.  If these assumptions are not valid, the radio
and infrared spectral types have to be seen as lower limits (i.e. the
ionising star is of earlier spectral type).

In column 4 of Table \ref{tab:sptype} the spectral type of the hottest
OB star found by our spectroscopic survey is given. It turns out that
in about 50 \% of the \HII\ regions we have identified the main
ionising source(s). This finding would suggest that the IRAS
fluxes of about 50 \% of the sources is dominated by the dust heated
by the cluster members instead of the emission of an UCHII region.
In some regions, however, the ionising sources are likely heavily
embedded and not visible in the near-infrared (e.g. \object{IRAS
11097-6102}).

\section{Discussion}\label{sec:discussion}

We have obtained $K$-band spectra of deeply embedded massive stars in
IRAS sources with UCHII colours. The stars are classified using the
classification scheme developed by \citet{Hanson96}. The question is
whether the properties of the young OB stars resemble those of the
more evolved OB stars in OB associations and the field
population. Also, do they  drive a normal stellar wind at such a young
age?  We will discuss the spectral properties of the stars as well as
age estimates of the clusters and the implications for the UCHII
lifetime problem.


\subsection{Stellar wind properties}\label{sec:indivstars}

The OB stars in this study are classified following
\citet{Hanson96}. In general these stars have $K$-band spectra similar
to field stars. To determine whether the properties of the young,
embedded OB stars are different from those in the field, the $K$-band
spectra of our sources are compared with $K$-band spectra of field
stars, like the spectra in \citet{Hanson96}. These are, however, also
used to classify the spectra.  This means that the lines used for the
classification cannot be used for this comparison. For example, in the
spectral classes kO9-B1 and later, the strength of \brg\ is used for
spectral classification and a different behaviour of \brg\ (i.e. due
to a stronger stellar wind) would result in an erroneous
classification. For the early types (kO3-O4 to kO7-O8), however, more
lines are present and \brg\ does not need to be used for the spectral
classification, so the properties of this line can be compared with
those of the field stars.

A few stars, classified as early O stars, show a peculiar behaviour of
the \brg\ line (Sect. \ref{sec:ourspectra}): the line is not present,
or in emission. Also the \heii\ line in one of the objects
(\object{16177nr405}) is either not present, or very weak. One of the
possible explanations for this effect is that the stars have a strong
stellar wind and that the photospheric \brg\ absorption is filled in
by stellar wind emission. This would only happen, however, if the wind
mass loss rate is as high as that of OB supergiants
\citep[cf.][]{Puls96,LenorzerModel04}. The \heii\ line is not formed
in the wind, but originates in the stellar photosphere.  This line
should be much less affected by the stellar wind than \brg. Only a
very high wind density makes the \heii\ line turn into emission. A
similar effect is found in the optical spectrum, where the \heii\ line
 at 4686 \AA\ turns into emission with increasing stellar
wind.  Veiling by dust or free-free emission of the surrounding \HII\
region seems unlikely as the intrinsically weak lines like \civ\ and
\niii\ are detected.

Contrary to the suggestion found in our data that the young stars
possess a relatively strong wind, evidence has been presented that in the
\object{SMC} a class of young OB stars (Vz stars) is found that
have unusually weak winds for their spectral type
\citep{Heydari02,Martins04}. The suggestion is that these stars are
located very close to or even on the ZAMS. The authors also suggest that
these objects might be sub-luminous.

 To disthinguish between these two scenarios for the stellar wind of
the youngest OB stars, a careful analysis of spectral lines sensitive
to the wind density is needed. For these deeply embedded stars, the UV
spectrum or the H$\alpha$ line which are usually employed to determine
the stellar wind properties are not available. The \brg\ line in the
$K$-band is not the best line for this ana\-ly\-sis. \bra\ in the
$L$-band is more sensitive to the stellar wind density
\citep{LenorzerModel04}.

\subsection{Binarity and stellar rotation}

The primordial binary fraction and the rotation rate of newly formed
massive stars carry important information regarding the star formation
process. Naively one would expect that newborn stars rotate more
rapidly than evolved stars. Furthermore, the fact that the majority of
massive stars are members of a binary/multiple system indicates that
many young massive stars should belong to binary systems as well.

The binarity of the OB stars detected in massive starforming regions
provides a strong constraint on formation
scenarios of massive stars \citep[e.g.][]{Zinnecker03}. Observations
of e.g. the massive stars in the Orion Nebula Cluster show that all
stars belong to double or multiple systems
\citep{Preibisch99}. Studies to determine the fraction of
spectroscopic binaries among bright O stars show that at least 35~\%
of the field O stars have a companion, while the fraction of
spectroscopic binaries in the case of young stellar clusters can get
as high as 60~\%
\citep{Mason98}. The contribution from single-line spectroscopic
binaries (SB1s) and double-line specroscopic binaries (SB2s) is
roughly equal.

The OB stars presented in this paper all belong to young stellar
clusters, suggesting that also among these massive stars a large
fraction is member of a binary system. With spectra taken at only one
epoch, we will not be able to detect the SB1s. To this aim we carried
out second-epoch observations of most of the OB stars in our sample
(Apai et al., in prep). For two out of 16 targets a shift in radial
velocity has been measured, making the targets SB1 candidates.  We
find no evidence for double-lined spectroscopic binaries in our
sample. Obviously, our spectra have limited spectral resolution and
signal to noise, and include only a few photospheric lines. Also, the
detectability of an SB2 depends on the ratio in luminosity between the two
components and the eccentricity of the system.

To illustrate the variation in projected rotational velocity in our
sample, the \hei\ lines at 2.1127 and 2.1138 $\mu$m are plotted
on a velocity scale for four different objects. In the top
spectrum the two
\hei\ lines are clearly separated, while in the bottom spectrum the two \hei\
lines are severely blended due to the rotational broadening. Proper
measures of the value of $v\sin{i}$ in these stars should be obtained
using more lines and higher spectral resolution, but at first
inspection the projected rotational velocities do not seem to be
different from those measured from main-sequence O-type stars
\citep{Penny96}.

\begin{figure}
 \includegraphics[width=0.8\columnwidth]{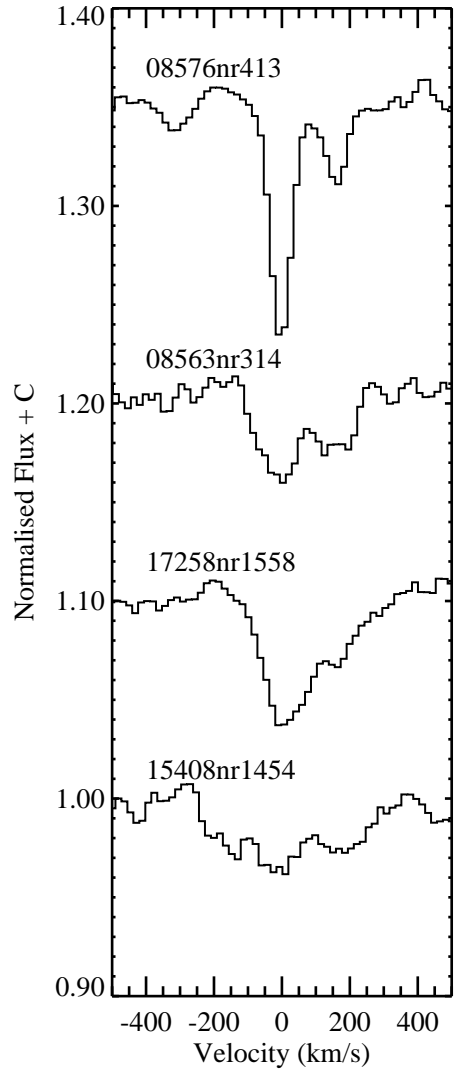}
   \caption{Profiles of the \hei\ lines at 2.1128 and 2.1138
   \micron. The different width of the lines reflects their rotation
   properties.  \label{fig:vsini}} 
\end{figure}

\subsection{Cluster age} \label{sec:starprop}

Despite the fact that most of the OB stars in our sample are not
directly associated with an UCHII region, their location deeply
embedded inside star forming regions (see Sect. \ref{sec:regions})
still suggests that these objects are young.  Studies of the embedded
population in \object{W31} \citep{Blum01} and \object{M17}
\citep{Hanson97} show that these regions are young and have an age of
$\sim 1$ million year. These age determinations were based on the
location of the stars in the HRD with respect to isochrones. However,
isochrone fitting is not reliable anymore for very young clusters
($\leq 1 - 2$~Myr) when only massive stars are used. The isochrones
are located too close to each other in the HR-diagram and the physical
parameters for the hottest stars are not well known. Comparing the
position of low-mass stars with the pre-main sequence tracks will
improve the age determination, as well as allow for a measurement of
the disk fraction in the young clusters \citep[e.g.][]{Lada03}.

\begin{figure}
 \includegraphics[width=\columnwidth]{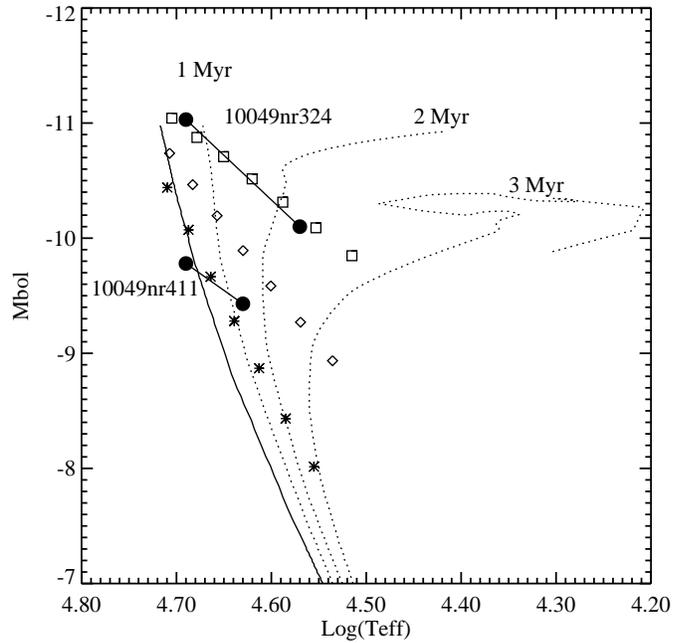}
   \caption{Upper part of the HR-diagram. The full line represents the
   ZAMS, the other lines are the isochrones of 1, 2 and 3 million
   years. The data are taken from \citet{Lejeune01}. The location of
   the two objects in IRAS 10049-5657 is represented by two bullets
   connected with a line. The end points of the lines correspond to
   the location of the earliest and latest spectral type consistent
   with the observed spectrum. The temperatures are taken from
   \citet{Martins02}, and the values for $V-K$ and the bolometric
   correction from \citet{Koornneef83} and \citet{LandoltBornstein},
   respectively.  Overplotted are the models for O stars taken from
   \citet{LenorzerModel04}. The crosses correspond to the position of
   the luminosity class V models for different T$_{\mathrm{eff}}$ and
   luminosity, while the diamonds and squares represent the locations
   of the giant and supergiant models, respectively.
 \label{fig:hrd}} 
\end{figure}

 The method of isochrone fitting is illustrated in
Fig. \ref{fig:hrd}. Two objects, \object{10049nr324} (O3-O6.5V) and
\object{10049nr411}(O3-O4V), are plotted in the HR-diagram together
with isochrones \citep{Lejeune01}.  The solid line is the ZAMS, the
other isochrones are those of 1, 2 and 3 million year. The two massive
stars located in \object{IRAS 10049-5356} (plotted with a
``$\bullet$'' in Fig. \ref{fig:hrd}) are located close to the upper
end of the ZAMS. Their location in the HR-diagram is consistent with
an age younger than $\sim 2$ million years. We note that the location
of \object{10049nr324} in the HR-diagram implies a luminosity of about
$3-4 \times 10^{6}\ \mathrm{L}_{\sun}$. This would make this object
one of the most luminous stars known in the galaxy comparable to the
enigmatic object $\eta$ Car (log(L/L$_{\sun}$) $\sim
6.5-6.7$). Remarkably, the \brg\ line of \object{10049nr324} shows one
of the strongest absorption profiles observed in any object classified
as O3 \citep{Hanson96}. However, at such extreme luminosity it is
unlikely that the radiation driven wind would be too weak to produce
\brg\ emission. This is confirmed by model calculations of \citet
{LenorzerModel04}, who predict that \brg\ should be in emission for
supergiants. For any combination of $L$ and $T_{\mathrm{eff}}$
consistent with the observations of \object{10049nr324}, \brg\ should
be in emission, or at least only weakly in absorption, contrary to
what is observed.

There are two possible solutions to this problem. The distance
estimate of the cluster may be in error. However, a distance of $\sim
3$~kpc would be required to be consistent with a position of
\object{10049nr324} on the ZAMS , which would move \object{10049nr411}
significantly below the ZAMS. Alternatively, \object{10049nr324} could
be a spectroscopic binary consisting of 2 equal mass O3-O4 stars. Both
solutions place one or both stars very close to the ZAMS, suggesting
an age of $\sim 1$~Myr or less.

\subsection{UCHII lifetime problem}

 A fundamental problem with UCHII regions is related to their
lifetime. If the young and ultra-compact H~{\sc ii} region expands due
to over-pressure with roughly the sound speed, the lifetime of an
UCHII is expected to be on the order of 10$^{4}$ years.  Using the
IRAS colour-colour criterium of \citet{WoodIRAS89}, the number of UCHII
regions in the galactic disk relative to the number of emerged O~stars
implies a duration of the embedded phase on the order of several
$10{^5}$ year, although a more refined evaluation shows that this
might be one order of magnitude less \citep{Comeron96}.


Several confining mechanisms have been proposed to extend the UCHII
lifetime: wind bow shocks generated by the supersonic motion of the
stars \citep[$v_{\rm sound} \simeq 0.2$~\kms,][]{vanBuren90},
evaporation of dense clumps or a circumstellar disk which mass-loads
the stellar wind \citep{Hollenbach94}, or an initial evolution taking
place in a very dense molecular cloud core.

\citet{Kaper05} show that in 75 \% of the observed fields a cluster is
identified towards the location of the IRAS source.  In
Sect. \ref{sec:regions} it is shown that in 50 \% of these clusters we
have detected the main ionizing source. In these cases the infrared
luminosity as measured by IRAS could be explained by the
spectroscopically identified OB stars in the clusters. This would mean
that in these cases the IRAS flux is not related to an UCHII radio
source but is caused by the OB stars located in the embedded cluster
heating the surrounding dust.

Estimates of the lifetime of these embedded clusters are of the order
of $10^{6}$ years \citep{Hanson97,Blum01}. As this lifetime is much
larger than those of the UCHII regions this would help to resolve the
``UCHII lifetime problem''. High spatial resolution observations in
the mid-infrared with e.g. Spitzer are required to test this
hypothesis.

\subsection{Ionising properties of the UCHII regions}

It is interesting to compare the different diagnostics concerning the
stellar content of the fields we studied, in relation to the spatial
scales they refer to.

The IRAS sources refer to scales of $\sim 1\arcmin$, comparable to
those of the embedded clusters. Similar scales apply to the single
dish radio measurements \citep{Caswell87}. As noted in Sect. \ref{sec:regions}, the
spectral types derived from both diagnostics are in reasonable
agreement, suggesting that on this scale we pick up optically thin
emission which is reprocessed light from the less embedded OB stars.


On small scales, we find radio emission from UCHII regions whose radio
continuum slope suggests that high-density, ionised gas is present.
We often find hot O stars as the ionising sources of the UCHII
regions. However, due to the optically thick nature of the radio
continuum, the contribution of these sources to the total radio
luminosity of the cluster is modest. Energy balance, however, requires
that the luminosity of these embedded O stars is re-emitted at long
wavelengths, i.e. far-infrared or mm. Since the IRAS flux can roughly
be accounted for by the reprocessed light from the less embedded
population, it may be that the reprocessed light of the more embedded
sources is emitted at wavelengths longwards of 100 \micron, outside
the IRAS wavelength range. This can be tested using e.g. HERSCHEL or
ALMA observations.

\section{Conclusions}\label{sec:conclusions}

In this paper, we present the $K$-band spectra of young OB stars deeply
embedded inside IRAS sources with UCHII colours. $K$-band
spectra of the actual near-infrared counterparts of the UCHII radio
sources are also presented and discussed. The results can be summarised as
follows:

\begin{itemize}
\item[$\bullet$]{38 OB type stars are identified inside massive
  star forming regions. They are classified based on the classification
  scheme of \citet{Hanson96}.
}
\item[$\bullet$]{The $K$-band spectra of the OB stars are in general
  similar to those of the (more evolved) OB stars located in OB
  associations and the field. The $K$-band spectral properties of the
  very young OB stars are already indistinguishable from those of the
  more evolved stars.}
\item[$\bullet$]{The spectra of a few O5-O6 stars suggest that these
  stars posses stronger winds than usually encountered in these
  stars. This finding needs to be confirmed by further observations of
  e.g. the \bra\ line.}
\item[$\bullet$]{Distances to the OB stars are derived using the
  ``spectroscopic parallax'' method. In 60 \% of the regions this
  distance is consistent with the kinematic  distance 
  or other determinations encountered in the literature. The distance
  determination towards the objects in the \object{Vela Molecular Ridge}
  suggest that \object{IRAS 08563-4711} and \object{09002-4732} are
  located in the Vela C cloud instead of the more distant Vela B cloud.}
\item[$\bullet$]{Comparison of the position of some of the OB stars with
  theoretical isochrones show that these regions are likely very young,
  less than 1-2 million years. This method to derive ages suffers from
  substantial uncertainties due to the  contribution from binaries and
  from fore- or background sources.}
\item[$\bullet$]{The differences in the line profiles of the \hei\
  2.112 and 2.113 lines reflect the difference in rotational
  broadening of the stars. At first sight, the projected rotational
  velocities of the young OB stars do not differ from those measured
  from main-sequence OB stars.}
\item[$\bullet$]{In most of the clusters (50\%), the ionising source
  is identified by means of our $K$-band spectroscopy. The $K$-band
  spectral type of the OB star(s) is in most cases similar to the
  spectral type derived from the infrared luminosity and the
  single-dish radio observations. This implies, in these
  cases, that the infared flux is produced by the embedded cluster and not by
  an associated UCHII region. This could help to resolve the UCHII
  lifetime problem as the embedded clusters have a lifetime of
  $10^{6}$ years compared to 10$^{4}$ years of the UCHII regions.}
\item[$\bullet$]{The $K$-band spectra of 7 point sources which are the
  near-infrared counterparts of UCHII radio sources show that the
  majority of these regions are dominated by nebular spectra. The
  \hei\ lines present in 6 of these objects suggest that O stars are
  located inside these regions.}
\item[$\bullet$]{Despite the fact that these UCHII regions include O
  stars, these stars are not dominating the single-dish radio
  flux as their \HII\ regions are still partially optically thick. The
  single-dish radio observations are dominated by the optically thin
  gas ionised by the OB stars in the cluster containing the UCHII region.}
\item[$\bullet$]{The infrared flux is also dominated by the dust
  heated by the stars in the cluster instead of the UCHII region.
  This suggests that the UCHII regions emit most of their flux at
  far-infrared and sub-mm wavelengths.}
\end{itemize}

\begin{acknowledgements}
AB acknowledges financial support from the DFG during a two-month
visit at ESO Headquaters. LK has been supported by a fellowship of the
Royal Academy of Arts and Sciences in the Netherlands. The authors
thank the VLT staff for support and help with the observations. Rens
Waters, Alex de Koter, Annique Lenorzer and Fernando Comer\'{o}n are
thanked for stimulating discussions. We thank Elena Puga-Antolin for
help with the identification of the near-infrared counterparts of the
UCHII regions. We thank the anonymous referee for his/her suggestions
to improve the paper. NSO/Kitt Peak FTS data used here were produced
by NSF/NOAO.  MMH acknowledges support from the National Science
Foundation, Grant No. 0094050.
\end{acknowledgements}


\appendix
\onecolumn
\section{Observing log}

\begin{center}
\tablefirsthead{\hline\hline 
 \multicolumn{1}{c}{IRAS}  &\multicolumn{1}{c}{nr}&\multicolumn{1}{c}{Class}&\multicolumn{1}{c}{Exp} &\multicolumn{1}{c}{S/N} &\multicolumn{1}{c}{Standard star} &\multicolumn{1}{c}{$K$}&\multicolumn{1}{c}{$J-K$}\\ \hline}

\tablehead{ \multicolumn{8}{l}{\small\sl continued from previous page}\\
\hline \hline  \multicolumn{1}{c}{IRAS}  &\multicolumn{1}{c}{nr}&\multicolumn{1}{c}{Class}&\multicolumn{1}{c}{Exp} &\multicolumn{1}{c}{S/N} &\multicolumn{1}{c}{Standard star} &\multicolumn{1}{c}{$K$}&\multicolumn{1}{c}{$J-K$}\\ \hline}

\tabletail{\hline \multicolumn{8}{r}{\small\sl continued on next page}\\ \hline}

\tablelasttail{\hline}

\bottomcaption{Identification of the OB stars and observing
  log. Column 1: associated IRAS pointsource, column 2: source number,
  taken from \citet{Kaper04}, column 3: class of object; OB: OB stars;
  YSO: candidate massive YSO; NEB: nebular UCHII radio source
  counterpart; LT: late-type stars, column 4: total integration time,
  column 5: obtained signal-to-noise ratio, column 6: observed
  standard star used to remove the telluric lines, columns 7 and 8:
  $K$ and $J-K$ values taken from \citet{Kaper04} $^\dagger$: This
  source is slightly extended and no pointsource photometry was
  possible.\label{tab:obslog}}
\begin{supertabular}{llllllll}
\object{06058$+$2138}&   221  &	 YSO	& 600    & 122 & \object{HD 43726} (A1V)   &  10.7$\pm$ 0.03  &  3.8$\pm$0.11  	\\ 
\object{06058$+$2138}&   227  &	 YSO	& 600    & 148 & \object{HD 43726} (A1V)   &  10.7$\pm$ 0.03  &  $>$7.0        	\\ 
\object{06058$+$2138}&   534  &	 OB	& 600    & 125 & \object{HD 43726} (A1V)   &  10.4$\pm$ 0.02  &  1.3$\pm$0.04   	\\ 
\object{06061$+$2151}&   153  &  LT     & 960    & 110 & \object{HD 43726} (A1V)   &  11.69$\pm$ 0.04 &  4.64$\pm$0.26      \\
\object{06061$+$2151}&   674A &	 OB	& 960    & 161 & \object{HD 43726} (A1V)   &  11.8$\pm$0.04   &  2.0$\pm$0.09   	\\
\object{06061$+$2151}&   676  &	 YSO	& 960    &  30 & \object{HD 43726} (A1V)   &  13.3$\pm$0.08   &  4.2$\pm$0.5   	\\
\object{06084$-$0611}&   114  &  YSO    & 960    & 200 & \object{HD 43726} (A1V)   &   9.3$\pm$0.01   &  4.7$\pm$0.08        \\       
\object{06084$-$0611}&   118  &  YSO    & 960    & 184 & \object{HD 43726} (A1V)   &  10.8$\pm$0.03   &  4.2$\pm$0.14        \\       
\object{06412$-$0105}&    94  &  LT     & 960    & 200 & \object{HD 48829} (B9V)   &  10.09$\pm$0.02  &  0.95$\pm$0.03      \\
\object{06412$-$0105}&   121  &  YSO    & 960    & 108 & \object{HD 48829} (B9V)   &  10.2$\pm$0.02   &  1.8$\pm$0.04        \\       
\object{06567$-$0355}&  122   &  LT     & 960    & 170 & \object{HD 48829} (B9V)   &  11.22$\pm$0.03  &  4.38$\pm$0.17       \\
\object{06567$-$0355}&  257   &  LT     & 960    & 140 & \object{HD 48829} (B9V)   &  13.00$\pm$0.07  &  1.74$\pm$0.14        \\
\object{06567$-$0355}&	557   &	 OB	& 960    & 141 & \object{HD 48829} (B9V)   &  11.6$\pm$0.04   &  2.6$\pm$0.10   	\\
\object{06567$-$0355}&	589   &	 OB	& 960    & 156 & \object{HD 48829} (B9V)   &  10.4$\pm$0.02   &  3.2$\pm$0.07   	\\
\object{07299$-$1651}&   43   &  YSO    & 960    & 145 & \object{HD 65674} (A0V)   &   9.4$\pm$0.01   &  4.7$\pm$0.09        \\       
\object{07299$-$1651}&  314   &  YSO    & 960    &  90 & \object{HD 65674} (A0V)   &  11.5$\pm$0.03   &   $>$6.0             \\       
\object{07299$-$1651}&  533   &  LT     & 960    & 170 & \object{HD 65674} (A0V)   &  11.6$\pm$0.04   &  3.34$\pm$0.13       \\       
\object{07299$-$1651}&  598   &  YSO    & 960    & 131 & \object{HD 65674} (A0V)   &  11.2$\pm$0.03   &  1.6$\pm$0.06        \\       
\object{07299$-$1651}&  618   &  YSO    & 960    & 175 & \object{HD 65674} (A0V)   &   9.4$\pm$0.01   &  4.8$\pm$0.09        \\       
\object{08563$-$4711}& 	314   &	 OB	& 600    & 189 & \object{HD 80998} (B8V)   &  8.9 $\pm$0.01   &  1.4$\pm$0.02   	\\
\object{08563$-$4711}& 	317   &	 OB	& 600    & 161 & \object{HD 80998} (B8V)   &  10.0 $\pm$0.02  &  2.3$\pm$0.04   	\\
\object{08576$-$4334}&  179   &  OB  	& 600    & 134 & \object{HD 80055} (A0V)   &  9.6$\pm$0.01    &  1.9$\pm$0.03   	\\
\object{08576$-$4334}&  201   &  LT  	& 600    & 115 & \object{HD 80055} (A0V)   &  9.61$\pm$0.02   &  2.70$\pm$0.04           \\
\object{08576$-$4334}&  225   &  LT  	& 600    & 230 & \object{HD 80055} (A0V)   &  9.45$\pm$0.01   &  3.26$\pm$0.05           \\
\object{08576$-$4334}&  292   &  YSO    & 960    & 107 & \object{HD 80055} (A0V)   &   9.4$\pm$0.01   &  2.4$\pm$0.03        \\       
\object{08576$-$4334}&  408   &  YSO    & 600    & 173 & \object{HD 80055} (A0V)   &   7.3$\pm$0.03   &  3.1$\pm$0.02         \\
\object{08576$-$4334}&	413   &  OB 	& 600    & 255 & \object{HD 80055} (A0V)   &  7.5$\pm$0.01    &  1.8$\pm$0.01   	\\
\object{08576$-$4334}&	461   &  OB 	& 600    & 63  & \object{HD 80055} (A0V)   &  12.7$\pm$0.06   &  1.7$\pm$0.11   	\\
\object{08576$-$4334}&	462   &  OB 	& 600    & 183 & \object{HD 80055} (A0V)   &  7.0$\pm$0.01    &  1.8$\pm$0.01   	\\
\object{09002$-$4732}&	697   &	 OB	& 960    & 160 & \object{HD 80998} (B8V)   &  10.7$\pm$0.03   &  5.4$\pm$0.22   	\\
\object{09002$-$4732}&	738  &	 LT	& 960    & 30  & \object{HD 80998} (B8V)   &  13.72$\pm$0.11  &  3.35$\pm$0.37       	\\
\object{10049$-$5657}&	261   &  LT     & 600    & 120 & \object{HD 91373} (A0V)   &  9.33$\pm$0.01   &  2.11$\pm$0.03   	\\
\object{10049$-$5657}&	324   &  OB     & 600    & 109 & \object{HD 91373} (A0V)   &  10.5$\pm$0.02   &  2.8$\pm$0.06   	\\
\object{10049$-$5657}&	411   &  OB 	& 600    &  84 & \object{HD 91373} (A0V)   &  11.7$\pm$0.04   &  2.9$\pm$0.11   	\\
\object{11097$-$6102}&	693   &	 YSO	& 600    & 160 & \object{HD 102152} (A0V)  &   9.6$\pm$0.01   &  3.4$\pm$0.04   	\\
\object{11097$-$6102}&	1122  &	 OB 	& 960    & 179 & \object{HD 102152} (A0V)  &  10.5$\pm$0.02   &  2.4$\pm$0.05   	\\
\object{11097$-$6102}&	1218  &	 YSO	& 600    & 183 & \object{HD 102152} (A0V)  &   8.4$\pm$0.01   &  5.6$\pm$0.09   	\\
\object{12073$-$6233}& 	1016  &  LT 	& 1200   & 134 & \object{HD 110062} (B9V)  &  7.96$\pm$0.01   &  8.89$\pm$0.01   	\\
\object{12073$-$6233}& 	1974  &  OB 	& 1200   & 168 & \object{HD 110062} (B9V)  &  10.7$\pm$0.03   &  2.3$\pm$0.05   	\\
\object{12073$-$6233}& 	2851  &  OB 	& 1200   &  96 & \object{HD 110062} (B9V)  &  10.7$\pm$0.03   &  2.5$\pm$0.06   	\\
\object{15408$-$5356}&	1410  &  OB 	& 600    & 140 & \object{HD 137251} (A2V)  &  8.6$\pm$0.01    &  2.2$\pm$0.02   	\\
\object{15408$-$5356}&	1454  &  OB	& 600    & 158 & \object{HD 137251} (A2V)  &  9.3$\pm$0.01    &  2.1$\pm$0.03   	\\
\object{15411$-$5352}&	462   &	 LT	& 600    & 100 & \object{HD 137251} (A2V)  &   8.14$\pm$0.01   &  6.95$\pm$0.13   	\\
\object{15411$-$5352}&	1955  &	 YSO	& 600    & 188 & \object{HD 137251} (A2V)  &   7.9$\pm$0.01   &  4.3$\pm$0.03   	\\
\object{15502$-$5302}&	2960  &	 OB	& 960    & 62  & \object{HD 137251} (A2V)  &  13.0$\pm$0.08   &  2.6$\pm$0.19   	\\
\object{15502$-$5302}&	3167  &	 OB 	& 960    & 106 & \object{HD 137251} (A2V)  &  11.8$\pm$0.04   &  1.6$\pm$0.19   	\\
\object{16128$-$5109}&  1182  &  LT     & 960    & 180 & \object{HD 150628} (A0V)  &  10.14$\pm$0.02  &  5.88$\pm$0.20          \\
\object{16164$-$5046}&	1542  &	 OB 	& 960    & 119 & \object{HD 150628} (A0V)  &  11.7$\pm$0.04   &  2.0$\pm$0.08   	\\
\object{16164$-$5046}&	3636  &	 YSO	& 960    & 137 & \object{HD 150628} (A0V)  &   9.5$\pm$0.02   &   $\ge$8.1           \\
\object{16177$-$5018}& 	271   &  OB 	& 960    &  40 & \object{HD 150628} (A0V)  &  11.8$\pm$0.04   &  3.4$\pm$0.14   	\\
\object{16177$-$5018}&	405   &  OB 	& 960    & 131 & \object{HD 150628} (A0V)  &  10.8$\pm$0.03   &  $\ge$6.7 	     	\\
\object{16177$-$5018}& 	1020  &  OB     & 960    &  98 & \object{HD 150628} (A0V)  &  11.9$\pm$0.04   &  4.7$\pm$0.27   	\\
\object{16177$-$5018}& 	2239  &  LT     & 960    &  60 & \object{HD 150628} (A0V)  &  8.77$\pm$0.01   &  2.53$\pm$0.02   	\\
\object{16571$-$4029}&	820   &  OB 	& 960    & 124 & \object{HD 158684} (A0V)  &   9.3$\pm$0.01   &  2.6$\pm$0.03   	\\
\object{16571$-$4029}&	1281  &  YSO  	& 960    & 132 & \object{HD 158684} (A0V)  &  9.3$\pm$0.01    &  2.5$\pm$0.06  	\\
\object{16571$-$4029}&	1610  &  OB  	& 960    & 132 & \object{HD 158684} (A0V)  &  11.0$\pm$0.03   &  2.5$\pm$0.07   	\\
\object{17136$-$3617}&	447   &  LT  	& 960    & 100 & \object{HD 161759} (A0V)  &  9.45$\pm$0.01    & 7.81$\pm$0.40  	\\
\object{17136$-$3617}&	649   &  YSO  	& 960    & 191 & \object{HD 161759} (A0V)  &  9.5$\pm$0.01    &  3.8$\pm$0.06  	\\
\object{17136$-$3617}&	712   &  LT  	& 960    & 170 & \object{HD 161759} (A0V)  &  8.55$\pm$0.01    & 6.05$\pm$0.11  	\\
\object{17149$-$3916}&	792   &  LT 	& 600    &  70 & \object{HD 161759} (A0V)  &  8.68$\pm$0.01   &  3.39$\pm$0.01   	\\
\object{17149$-$3916}&	895   &  OB 	& 600    & 139 & \object{HD 161759} (A0V)  &  8.3 $\pm$0.01   &  0.6$\pm$0.01   	\\
\object{17149$-$3916}&	1061  &  LT 	& 600    &  30 & \object{HD 161759} (A0V)  &  10.42 $\pm$0.02   &  5.47$\pm$0.17   	\\
\object{17149$-$3916}&	3361  &  LT 	& 600    &  50 & \object{HD 161759} (A0V)  &  12.87 $\pm$0.06   &  1.50$\pm$0.10   	\\
\object{17175$-$3544}&  59    &  OB     & 960    & 166 & \object{HD 161759} (AOV)  &  11.5$\pm$0.03   &  6.3$\pm$0.5         \\
\object{17258$-$3637}&	378   &	 OB	& 960    & 141 & \object{HD 161759} (A0V)  &  10.2$\pm$0.02   &  3.4$\pm$0.06   	\\
\object{17258$-$3637}&	593   &	 YSO	& 960    & 200 & \object{HD 161759} (A0V)  &   9.6$\pm$0.01   &  5.5$\pm$0.17   	\\
\object{17258$-$3637}&	1558  &  OB     & 960    &  17 & \object{HD 161759} (A0V)  &  10.7$\pm$0.02   &  1.9$\pm$0.04   	\\
\object{17423$-$2855}&	3102  &	 OB	& 960    & 113 & \object{HD 166131} (B8V)  &  9.9$\pm$0.02    &  5.3$\pm$0.13   	\\
\object{17423$-$2855}&	3130  &	 LT	& 960    &  60 & \object{HD 166131} (B8V)  & 10.81$\pm$0.03    &      $>$6.8   	\\
\object{17423$-$2855}&	6909  &	 LT	& 960    &  80 & \object{HD 166131} (B8V)  &  9.47$\pm$0.01    &  15.89$\pm$0.18   	\\
\object{17574$-$2403}&  930   &  LT     & 960    &  80 & \object{HD 171572} (B9V)  &  9.89$\pm$0.02  &    $>$ 7.8       \\
\object{17574$-$2403}&  3471  &  LT     & 960    & 100 & \object{HD 171572} (B9V)  &  10.53$\pm$0.02  &  1.27$\pm$0.04       \\
\object{17574$-$2403}&  xx$^\dagger$    &  NEB    & 960    &  32 & \object{HD 171572} (B9V)  &   --             &  --                     \\
\object{18006$-$2422}&   766  &  YSO	& 600    & 146 & \object{HD 171808} (A0V)  &  9.4 $\pm$0.01   &  2.8$\pm$0.04   	\\
\object{18006$-$2422}&   770  &  OB 	& 600    & 131 & \object{HD 171808} (A0V)  &  7.3 $\pm$0.01   &  0.9$\pm$0.01   	\\
\object{18006$-$2422}&   702  &  LT 	& 600    & 102 & \object{HD 171808} (A0V)  &  9.93 $\pm$0.02   & 4.29 $\pm$0.08   	\\
\object{18006$-$2422}&  1134  &  LT 	& 600    & 150 & \object{HD 171808} (A0V)  &  9.04 $\pm$0.04   & 3.49 $\pm$0.04   	\\
\object{18006$-$2422}&   2035 &  NEB 	& 600    &  88 & \object{HD 171808} (A0V)  &  11.4 $\pm$0.04   & 1.8 $\pm$0.07   	\\
\object{18060$-$2005}&	719   &  LT 	& 600    &  70 & \object{HD 171808} (A0V)  &  9.49$\pm$ 0.01  &  5.54$\pm$0.13   	\\
\object{18060$-$2005}&	1073   &  LT 	& 600    &  80 & \object{HD 171808} (A0V)  & 10.46$\pm$ 0.02  &  5.26$\pm$0.18   	\\
\object{18060$-$2005}&	1733  &  OB 	& 600    & 100 & \object{HD 171808} (A0V)  &  9.0$\pm$ 0.01   &  2.9$\pm$0.03   	\\
\object{18060$-$2005}&	2481  &	 OB 	& 960    & 75  & \object{HD 171808} (A0V)  &  11.1$\pm$0.03   &  3.5$\pm$0.11   	\\
\object{18449$-$0158}&	319   &  OB 	& 960    &  93 & \object{HD 181690} (B9V)  &  10.8$\pm$0.03   &  2.4$\pm$0.06   	\\
\object{18449$-$0158}&	335   &  OB 	& 960    & 159 & \object{HD 181690} (B9V)  &  8.5$\pm$0.01    &  7.0$\pm$0.17   	\\
\object{18449$-$0158}&	1493   &  LT 	& 960    &  80 & \object{HD 181690} (B9V)  &  10.71$\pm$0.03  &  2.37$\pm$0.06   	\\
\object{18507$+$0110}&  262   &  OB 	& 960    & 190 & \object{HD 181690} (B9V)  &  9.4$\pm$0.01    &  4.3 $\pm$0.07  	  \\
\object{18507$+$0110}&  248   &  YSO 	& 960    & 153 & \object{HD 181690} (B9V)  & 10.3$\pm$0.02    &  5.4 $\pm$0.18  	  \\
\object{18507$+$0110}&	373   &  OB     & 960    &  70 & \object{HD 181690} (B9V)  &  12.3$\pm$0.05   &  5.4$\pm$0.49        \\
\object{18507$+$0110}&	389   &  OB 	& 960    &  90 & \object{HD 181690} (B9V)  &  11.3$\pm$0.03   &  3.8 $\pm$0.14  	\\
\object{18592$+$0108}&	1035  & LT 	& 960    & 120 & \object{HD 181690} (B9V)  &  11.31$\pm$0.03   & 4.08 $\pm$0.14    \\
\object{18592$+$0108}&	1051  & LT 	& 960    & 130 & \object{HD 181690} (B9V)  &  11.29$\pm$0.03   & 5.00 $\pm$0.21    \\
\object{18592$+$0108}&	1555  & NEB 	& 960    &  40 & \object{HD 181690} (B9V)  &  12.6$\pm$0.05   &  $>$ 5.1    	\\
\object{19078$+$0901}&  185   &  LT     & 960    &  90 & \object{HD 186549} (B9V)  &  10.69$\pm$0.02   &  3.74 $\pm$0.10        \\
\object{19078$+$0901}&  438   &  LT     & 960    & 110 & \object{HD 186549} (B9V)  &  11.63$\pm$0.04   &  4.10 $\pm$0.19        \\
\object{19078$+$0901}&  549   &  LT     & 960    & 130 & \object{HD 186549} (B9V)  &  11.88$\pm$0.04   &  5.01 $\pm$0.33        \\
\object{19078$+$0901}&  647   &  OB     & 960    &  65 & \object{HD 186549} (B9V)  &  13.2$\pm$0.07   &  4.4 $\pm$0.5        \\
\object{19078$+$0901}&  670   &  LT     & 960    &  80 & \object{HD 186549} (B9V)  &  10.11$\pm$0.02   &  2.17 $\pm$0.04        \\
\object{19078$+$0901}&  1202  &  LT     & 960    & 120 & \object{HD 186549} (B9V)  &  10.31$\pm$0.02   &  3.75 $\pm$0.09       \\
\object{19078$+$0901}&  2070  & NEB     & 960    &  32 & \object{HD 186549} (B9V)  &  13.5$\pm$0.07   &  $>$ 4.2        \\
\object{19078$+$0901}&  2224  & NEB     & 960    &  33 & \object{HD 186549} (B9V)  &  13.1$\pm$0.07   &  $>$ 4.6        \\
\object{19111$+$1048}&  207   & LT      & 960    &  70 & \object{HD 186549} (B9V)  &  11.23$\pm$0.05   &  2.77$\pm$0.12       \\ 
\object{19111$+$1048}&  1149  & LT      & 960    &  70 & \object{HD 186549} (B9V)  &  11.15$\pm$0.05   &  1.89$\pm$0.08       \\ \hline

\end{supertabular}
\end{center}


%
%


\end{document}